\documentclass{emulateapj}

\newcommand{\hd}{H$\delta$}%
\newcommand{\hg}{H$\gamma$}%
\newcommand{\hb}{H$\beta$}%
\newcommand{\hal}{H$\alpha$}%
\newcommand{\lya}{Ly$\alpha$}%
\newcommand{\SII}{[S{\sc ii}]}%
\newcommand{\OI}{[O{\sc i}]$\lambda6300$}%
\newcommand{\OIIIA}{[O{\sc iii}]}%
\newcommand{\OIIID}{[O{\sc iii}]$\lambda\lambda4959,5007$}%
\newcommand{\OIIIE}{[O{\sc iii}]$\lambda4959$}%
\newcommand{\OIIIF}{[O{\sc iii}]$\lambda5007$}%
\newcommand{\OIIIG}{[O{\sc iii}]$\lambda4363$}%
\newcommand{\OII}{[O{\sc ii}]$\lambda3727$}%
\newcommand{\NII}{[N{\sc ii}]}%
\newcommand{\CIV}{[C{\sc iv}]}%
\newcommand{\NeIII}{[Ne{\sc iii}]}%
\newcommand{\NeV}{[Ne{\sc v}]}%
\newcommand{\kmsec}{km\,sec$^{-1}$}%
\newcommand{\ergsec}{erg\,sec$^{-1}$}%

\def\gs{\mathrel{\raise0.35ex\hbox{$\scriptstyle >$}\kern-0.6em
\lower0.40ex\hbox{{$\scriptstyle \sim$}}}}
\def\ls{\mathrel{\raise0.35ex\hbox{$\scriptstyle <$}\kern-0.6em
\lower0.40ex\hbox{{$\scriptstyle \sim$}}}}

\newcommand{\msun}{{\rm\,M_\odot}}

\newcommand{\lsun}{\,{\rm L}_\odot}

\def\ltsima{$\; \buildrel < \over \sim \;$}
\def\simlt{\lower.5ex\hbox{\ltsima}}
\def\gtsima{$\; \buildrel > \over \sim \;$}
\def\simgt{\lower.5ex\hbox{\gtsima}}

\shorttitle{Spectral Properties of SMGs}
\shortauthors{Takata et al.}

\begin{document}

\title{Restframe Optical Spectroscopic Classifications for Submillimeter Galaxies\altaffilmark{8}\altaffilmark{9}\altaffilmark{10}}

\author{Tadafumi Takata,\altaffilmark{1,2} Kazuhiro
  Sekiguchi,\altaffilmark{3} Ian Smail,\altaffilmark{4} Scott C.\
  Chapman,\altaffilmark{5} J.\,E.\ Geach,\altaffilmark{4} A.\,M.\
  Swinbank,\altaffilmark{4} Andrew Blain,\altaffilmark{5} \& R.\,J.\
  Ivison\altaffilmark{6,7}}

\altaffiltext{1}{Astronomy Data Center, National Astronomical Observatory of Japan, 2-21-1 Osawa, Mitaka, Tokyo 181-8588, Japan}
\altaffiltext{2}{Email: tadafumi.takata@nao.ac.jp}
\altaffiltext{3}{Subaru Telescope, NAOJ, 650 North A'ohoku Place, Hilo, HI 96720}
\altaffiltext{4}{Institute for Computational Cosmology, Department of Physics, 
       Durham University, South Road, Durham DH1 3LE, UK}
\altaffiltext{5}{Astronomy Department, California Institute of Technology, 105-24, Pasadena, CA 91125}
\altaffiltext{6}{Astronomy Technology Center, Royal Observatory, Blackford Hill, Edinburgh EH9 3HJ, UK}
\altaffiltext{7}{Institute for Astronomy, University of Edinburgh, Blackford
  Hill, Edinburgh EH9 3HJ, UK}

\altaffiltext{8}{Based partly on data collected at Subaru Telescope and obtained from the 
SMOKA science archive at Astronomical Data Analysis Center, which are operated by the National 
Astronomical Observatory of Japan.}
\altaffiltext{9}{Based partly on the data collected with the ESO VLT-UT1 Antu Telescope(074.B-0107(A)).}
\altaffiltext{10}{Based partly on data obtained at the Keck Observatory, which is operated as a 
scientific partnership among the California Institute of Technology, the University of California, 
and the National Aeronautics and Space Administration. The observatory was made possible by the 
generous financial support of the W.\,M.\ Keck Foundation. }

\begin{abstract}
  We report the results of a systematic near-infrared spectroscopic
  survey using
  the Subaru, VLT and Keck Telescopes 
 of a sample of high redshift Ultra-luminous Infrared Galaxies
  (ULIRGs) mainly composed of submillimeter-selected galaxies.  
Our observations span the restframe optical range containing nebular emission lines such as
  \hb, \OIIID, and \OII, which are essential for making robust 
  diagnostics of the physical properties of these ULIRGs.  Using the
  \hal/\hb\ emission line ratios, we derive internal extinction
  estimates for these galaxies similar to those of local ULIRGs:
$A_V\sim 2.9\pm 0.5$.  
Correcting the H$\alpha$ estimates of the star formation rate
  for dust extinction using the Balmer decrement, results in rates
  which are consistent with those estimated from the far-infrared luminosity.
  The majority ($>60$\%) of our sample show  spectral features
  characteristic of AGN (although we note this partially reflects an
  observational bias in our sample), with $\sim65$\% exhibiting broad
  Balmer emission lines. A proportion of these sources show relatively low
  \OIIIF/\hb\ line ratios, which are similar to those of Narrow Line
  Seyfert 1 galaxies suggesting small mass black holes which are rapidly 
  growing. 
 In the subsample of our survey with both
\OIIIF\ and hard X-ray coverage, at
  least $\sim 60$\% show 
  an excess of \OIIIF\ emission, by a factor of 5--10$\times$, relative to the hard X-ray
  luminosity compared to the correlation between these two properties
  seen in Seyferts and QSOs locally.  From our spectral diagnostics, we
  propose that the strong \OIIIF\
  emission in these galaxies arises from shocks in dense gaseous
  regions in this vigorously star-forming population.
 We caution that due to sensitivity and resolution limits, our
  sample is biased to strong line emitters and hence our
results do not yet provide a complete view of the 
physical properties of the whole high-redshift ULIRG population.    
\end{abstract}


\section{Introduction} 

There is almost irrefutible evidence for an increase in the star
formation density with redshift, as demonstrated by emission line and
continuum star formations tracers in wavebands from the ultraviolet to
the submillimeter and radio wavebands.  This evolution appears to be
stronger for tracers which are less sensitive to dust obscuration
(e.g.\ Ivison et al.\ 2006), suggesting that an increasing proportion
of the activity in more distant galaxies may be highly
obscured (e.g.\ Blain et al.\ 1999, 2002).  Indeed, 
recent results on the mid- to far-infrared emission of luminous
but dust obscured 
galaxies at high redshift ($z\sim1$--3) suggests that the 
origin of their large infrared luminosities is a mix of dust obscured 
vigorous star formation and/or dust enshrouded active galactic nucleus (AGN) 
(Yan et al.\ 2005; Houck et al.\ 2005; Lutz et al.\ 2005; Desai et al.\
2006). In many sources it is likely that both AGN and star formation
contribute to the emission as a result of the 
 close link required between the growth of 
super-massive black holes and bulges in massive galaxies (e.g.\ Borys et al.\
2005).

One of the best-studied populations of high-redshift, far-infrared
luminous galaxies is that identified in the submillimeter waveband
using the SCUBA camera (Holland et al.\ 1999) on the James Clerk
Maxwell Telescope (JCMT). Although they span less than an order of
magnitude in submillimeter flux, these galaxies are responsible for
much of the energy density in the submillimeter background (Barger et
al.\ 1998; Hughes et al.\ 1998; Smail et al.\ 2002; Cowie, Barger \&
Kneib 2002; Scott et al.\ 2002).  The faintness of these obscured
galaxies in the optical waveband has made it difficult to obtain
precise redshifts (e.g.\ Simpson et al.\ 2004), although some
progress has been made using ultraviolet/blue spectrographs (Chapman et
al.\ 2003a; 2005).  The median redshift for submillimeter galaxies with
850\,$\mu$m fluxes of $\gs 5$\,mJy, (hereafter SMGs) is $<\! z\! >\sim
2.2$ (Chapman et al.\ 2003a, 2005). The submillimeter and radio fluxes
of these systems indicate their bolometric luminosities are $\gs
10^{12}$\,L$_\odot$ (Kovacs et al.\ 2006), confirming that they are
examples of high-redshift Ultraluminous Infrared Galaxies (ULIRGs).
  
This population provides critical constraints on models of galaxy
formation and evolution. In particular, if the bolometric emission from
SMGs is powered solely by star formation, then these galaxies form
about half of the stars seen in the local Universe (Lilly et al.\
1999). However, it appears likely that both AGN and star formation
activity contribute to the immense far-infrared luminosities of these
systems, although it has been difficult to disentangle the precise
balance between these two energy sources.  Recent sensitive X-ray
analysis suggest that star formation is likely to be the dominant
source of the bolometric luminosity in SMGs (Alexander et al.\
2005a,b).  Further evidence suggest it is plausible to identify SMGs as
the progenitor of massive elliptical galaxies at the present-day, based
on their large gas, stellar and dynamical masses (Neri et al.\ 2003;
Greve et al.\ 2005; Tacconi et al.\ 2006; Smail et al.\ 2004; 
Borys et al.\ 2005; Swinbank et al.\ 2004, 2006).  
Furthermore, combining the X-ray constraints on the
AGN within this population with the typical mass estimates suggests
that SMGs are the sites of coeval growth of stellar bulges and central
black holes (Borys et al.\ 2005).

Rest-frame optical emission lines provide a powerful tool to
investigate many fundamental properties of galaxies, such as star
formation rates (SFRs), power sources, internal extinction and
metallicity.  Swinbank et al.\ (2004) conducted a systematic
near-infrared spectroscopic survey of thirty SMGs to investigate their
SFRs and metallicities and the kinematics of the emission line
gas. However, the wavelength coverage was limited to the region around
H$\alpha$ and so they did not include several emission lines at shorter
wavelengths, such as \hb\ and \OIIID, which are useful for evaluating
internal extinction and metallicity or determining the power source.

We present in this paper the results from a near-infrared spectroscopic
survey of redshifted \OIIID, \hb\ and \OII\ lines for a sample of
far-infrared luminous galaxies. The sample is composed of SMGs and
optically faint radio galaxies (OFRGs), at $z\sim1$--3.5.  Chapman et
al.\ (2004) and Blain et al.\ (2004) claim that high-redshift OFRGs are
ULIRGs, with similar bolometric luminosities to SMGs but warmer
characteristic dust temperature, resulting in them being undetectable
in the submillimeter waveband.  We use \hal/\hb\ emission line ratios
to derive the dust extinction in these systems and then employ these
estimates to derive extinction-corrected SFRs from the \hal\
luminosities. In addition, we also use X-ray observations of these
objects to compare the strength of the \OIIIF\ emission to their X-ray
emission, and so investigate the power of the AGN in these galaxies.
We adopt cosmological parameters of H$_0=$72 km sec$^{-1}$ Mpc$^{-1}$,
and $\Omega_M=0.3$ and $\Omega_{\Lambda}=$0.7 throughout.

\section{Observation and Data Reduction} 

Our sample was selected from the catalogs of SMGs and OFRGs in Chapman
et al.\ (2005, 2004).  We chose
SMGs/OFRGs in the redshift ranges $z=2.05$--2.56 and $z=1.28$--1.68, where
nebular emission lines such as [O{\sc ii}],
H$\beta$, [O{\sc iii}] and/or H$\alpha$ are redshifted into clear
parts of the $J$, $H$ and $K$-bands respectively.  
In total 22 targets were observed using the OHS
spectrograph on Subaru, ISAAC on the VLT or NIRSPEC on Keck. 
The log of the observations is given in Table~1.

\subsection{Subaru OHS observations and data reduction} 

The majority of our spectroscopic observations 
were taken with the OH Suppression
Spectrograph (OHS; Iwamuro et al.\  2001) with the Cooled Infrared
Spectrograph and Camera for OHS (CISCO; Motohara et al.\ 2002) attached
to the Nasmyth focus of Subaru Telescope (Iye et al.\ 2004).
Observations were obtained on the
nights of 2004 April 6, 7, June 24--25, and 2005 Feb 14--16.
Sky conditions were photometric on all these nights with typical seeing
0.5--0.7$''$ at 1.6$\mu$m.  We used a slit width of 0.95$''$, which gives a
resolution of $\Delta\lambda$/$\lambda\sim200$ ($\sim
1400$\,km\,sec$^{-1}$) and used the ``SP4'' dither pattern, which shifts
the object along the slit to four positions in one sequence.
After completing each observation, we observed bright A- or F-type stars
with the same configuration as the science observation to 
calibrate the extinction and sensitivity
variation with wavelength.
During each night we observed at least two photometric standard stars selected
from the UKIRT Faint Standards catalog (Hawarden et al.\ 2001). 
We used FS\,27 and FS\,127 for the observations taken in 2004 April, FS\,23
and FS\,30 in 2004 June and FS\,133 and FS\,127 in 2005 February.

The data reduction was performed in the standard manner using custom
scripts in {\sc iraf} and some {\sc c} programs provided by the OHS/CISCO
instrument teams. First, we subtracted the
sky background using the object frames
at different dither positions. Next we
fitted the  sky line residuals using two dimensional polynomials and
subtracted these from the data. We then shift-and-added
the images from the different dithering positions, using a median combine.
As the instrument is stable, wavelength calibration was performed using the nominal conversion of
pixel coordinates to wavelength.
To confirm the stability of the wavelength
solution, we analysed Argon calibration lamp
exposures taken during our runs 
and  checked for systematic shifts in wavelength.  We
found typical systematic shifts of  7--9\AA\ ($\sim 0.05$\%)
which is ignorable in our analysis due to the low
resolution of our spectra. 
Extinction, sensitivity and photometric calibration were
performed by dividing the calibrated spectra with those of the bright A- or
F-type standard star observations after fitting the stellar spectra
with models.

\subsection{VLT ISAAC observations and data reduction} 

We conducted
observations of four SMGs and one OFRG using the ISAAC spectrograph 
on the 8-m VLT on 2004 November 22--23 (Table~1). 
ISAAC was used in medium-resolution mode, which
provides spectral resolution of 3000 ($\sim100$ \kmsec). 
Seeing was steady at $\sim0.8''$ over the course
of the observations and the observations were taken with a
standard 10$''$ ABBA chop.  
Preliminary data reduction was performed using
the {\sc eclipse}\footnote{v4.9-0
  http://www.eso.org/projects/aot/eclipse} pipeline, using flat-fields
generated from night-calibrations taken after each observation, and
wavelength calibration from a solution using the OH sky lines.
The remaining flux-calibration was achieved in {\sc iraf}, using
corresponding Hipparcos standard stars observed throughout the
observing run and near-infrared fluxes derived from the 2MASS catalog.

\subsection{Keck NIRSPEC observations and data reduction} 

The observations of SMM\,J09431+4700 (H6/H7) and SMM\,J131201.17+424208.1 were
taken on 2004 April 8 in photometric conditions and 0.8$''$ seeing
using the NIRSPEC spectrograph on Keck.  These
observations employed the standard ABBA
configuration to achieve sky subtraction. Each exposure was 600\,s
in length and the total integration time was 2400\,s. The data were reduced
using the {\sc wmkonspec} package in {\sc iraf}.  We remapped the two
dimensional spectra using linear interpolation to rectify the spatial
and spectral dimensions. After subtracting pairs of nod-positions (the
nod was 20$''$ along the slit), residual sky features were removed in
{\sc idl} using sky regions on either side of the object spectrum.
For the wavelength calibration we used an Argon arc lamp.  The output
pixel scale is 4.3 \AA pix$^{-1}$, and the instrumental profile has a
FWHM of 15 \AA\ (measured from the widths of the sky-lines), which
corresponds to $\sim200$ \kmsec.  We used FS\,27 for photometric
calibration.

\section{Results}

\subsection{General Spectral Features} 

We show all of our
spectra in Figure~1. We identified emission lines in 20 spectra out of 22
targets which were observed.  Most of the SMGs show weak \hb\ emission,
but many show strong (and sometimes broad and distorted) profiles in
\OIIID.  Some of our spectra show additional
emission lines of \NeIII, \NeV and \OI, which are common in AGN.
Five of the SMGs from our sample (SMM\,J09431+4700 (H6), SMM\,J123549.44+621536.8, 
SMM\,J123716.01+620323.3, SMM\,J163639.01+405635.9, and SMM\,J163650.43+405734.5) 
display spatially extended structures ($\geq$1''.0) in either \OIIIF\ and/or \hal\ 
emission line (Figure~2, see Smail et al.\ 2003; Swinbank et al.\ 2005
for evidence of the spatial extension in
SMM\,J163650.43+405734.5). 

\subsection{Comments On Individual Objects}
Several of our observations are particularly noteworthy and we discuss them here.

\subsubsection{SMM\,J02399$-$0134}
This galaxy is identified as a submillimeter source associated with a
spiral galaxy at $z=1.06$, which shows features typical of a Seyfert 1 
(Smail et al.\ 1997, 2002; Soucail et al.\ 1999).  
The strong and featureless continuum, together with the spatially compact 
emission line flux indicates AGN activity; an interpretation which is further
supported by the detection of this source
in hard X-rays by Bautz et al.\ (2000). 
Our spectrum shows at least two peaks in the \hal\ emission line
with FWHM$_{\rm rest}\sim 200$--400\kmsec, consistent with these lines 
arising from independent components within the system.   
If we force fit a single Gaussian profile to
the \hal\ emission, we determine FWHM$_{\rm rest}=1530\pm500$\kmsec, 
which if it arises
from an AGN is narrower than typical Seyfert 1 galaxies, 
although broader than Seyfert 2 galaxies ($\sim500$\kmsec). 
This source is also detected by CO observation by Greve et al.\ (2005) 
with a double peaked profile with a FWHM of $780\pm60$\,\kmsec
and a separation  between the two peaks of $\sim400$\kmsec,
consistent within the errors with our measurements from \hal.
We therefore choose to interpret the double-peaked \hal\ line
as evidence for a merger or interaction in this system, with any
AGN-produced broad component undetected in our spectrum.

\subsubsection{SMM\,J09431+4700}

This source was discovered by Cowie, Barger \& Kneib (2002), and has
been identified with
two distinct $\mu$Jy radio counterparts: H6 and H7
(Ledlow et al.\ 2002).  
These are lensed sources, lying behind a massive cluster Abell\,851 at
$z=0.41$ although the amplification is modest: $1.3\times$. 
The redshift for H6 was measured by Ledlow et al.\ (2002) as $z=3.349$
from Ly$\alpha$, H7 was not observed.  The restframe ultraviolet
properties of H6 suggest it hosts an AGN with spectral
features similar to a narrow-line Seyfert 1 (Ledlow et al.\ 2002).
We placed the NIRSPEC slit across both radio components and detected \OIIIF\
emission from both sources at redshifts of $z=3.350$ and $z=3.347$ for
H6 and H7, respectively.  We also detected narrow (FWHM$_{rest}\sim350$
\kmsec) \hb\ emission from H6.  The \OIIIF\ emission from H6 is
spatially extended($>2''$ or 14.5\,kpc; Figure~2), but has no
significant velocity gradient across $\sim8$\,kpc in projection.  No hard
X-ray emission was detected with the upper limits on f$_{\rm 2-10KeV}$ as
$\sim 1\times$10$^{-15}$ erg sec$^{-1}$ cm$^{-2}$ 
(Ledlow et al.\ 2002). CO line emission is also detected by 
Neri et al.\ (2003) and Tacconi et al.\ (2006) based on our
restframe optical redshift, originating from H7 at  $z=3.346$. 
Millimeter continuum emission has been seen from H6, but assuming the
gas reservoir is at the redshift we find from \OIIIF, the gas mass
of the AGN-dominated component, H6, is a factor of a few lower than
that of H7.

\subsubsection{SMM\,J123549.44+621536.8}

This source has apparent double-peaked,  narrow ($<1500$\kmsec)
emission lines in \OII\ and \OIIID, with the two components
spatially offset by $\sim 0.2''$.  
The one dimensional spectra also shows signs of broad \hb\
emission at  $z=2.195\pm0.005$ with a FWHM of $2100\pm500$
\kmsec.  Both the \OIIIF\ and the \OII\ emissions are spatially
extended with faint wings on scales of approximately 1$''$
($\sim8$\,kpc), see Figure~2.  
There may also be a very weak, broad multiplet 
of Fe{\sc ii}$\lambda5190,5320$ (Figure~1), potentially indicating
the presence of the Narrow Line Seyfert 1 (NLS1) type AGN
component (Osterbrock \& Pogge 1985; Goodrich 1989).  This is
consistent with the results of Alexander et al.\ (2005b), which indicated
the presence of a heavily obscured AGN with N$_H \sim
10^{24}$\,cm$^{-2}$ based on their X-ray spectral analysis.
The spatial extension in the bright core of the \OIIIF\ likely
indicates merging components or rotation along the slit, while the extended
wings may reflect ``superwind'' activity. 

\subsubsection{SMM\,J123716.01+620323.3}

This source is very bright in the optical ($R_{AB} = 20.2$)
with a redshift of $z=2.053\pm0.005$ and it was classified as a QSO by
Chapman et al.\ (2005) based on the broad rest UV emission lines and 
comparable luminosities in rest optical and far-infrared 
wavelength, which exceed 
10$^{45}$ erg sec$^{-1}$. The source has also been detected in
hard X-rays by Alexander et al.\ (2005b).
Our spectrum shows several Hydrogen Balmer
lines such
as \hb, \hd\ and \hg\ with broad FWHM$_{rest}$ 
($\sim2200$--2700\,km\,sec$^{-1}$) and the \OIIID\ doublet with 
FWHM$_{rest}$ of $\sim2200$\,km\,sec$^{-1}$.  
We also detected the \NeIII\ and several Fe{\sc ii} lines at 
3--4$\sigma$ significance. The restframe 
optical spectrum is dominated by continuum emission 
without stellar absorption features, suggesting a large contribution 
from the AGN component to the total rest-frame optical flux. The 
\OIIIF\ emission lines are wide
FWHM$_{\rm rest} \sim2000$\kmsec\ and spatially
extended ($\sim1.5''$; 12\,kpc)   indicating 
dynamically active gas motion (Figure~2). 
The estimated Hydrogen column density from the X-ray 
spectral analysis is relatively low (N$_H \sim 10^{22.5}$\,cm$^{-2}$), 
which implies the AGN does not suffer from large extinction.  
It should be noted that the redshift based on the 
restframe-UV emission lines is 
$2.037\pm0.002$, which is blueshifted
by $1600\pm700$\kmsec\ from the  redshift indicated by the
restframe optical nebular emission line. This velocity offset may arise
due to broad \lya\ emission which may be affected by dust extinction
and resonance scattering.  

\subsubsection{SMM\,J131222.35+423814.1}

This source is another example of a NLS1 type AGN.
It lies at $z=2.560$ and our spectrum displays
broad \hb\ emission, with FWHM$_{\rm rest}\sim 2600\pm1000$\,\kmsec\ and 
a low \OIIIF/\hb\ ratio ($0.46^{+0.35}_{-0.28}$). This source has \lya\, \CIV 
and He{\sc ii} emission lines in the rest UV spectrum and 
was classified as a QSO by 
Chapman et al.\ (2005). The rest-frame optical emission is dominated by
very strong continuum emission without stellar absorption lines, supporting 
the presence of a luminous AGN component. Unfortunately, there is no coverage 
of \hal\ emission for this object and so we could not constrain the internal 
extinction. The \NeV\ line 
(which is a very clean indicator of AGN activity; Osterbrock 1989) 
is detected. Furthermore, this source was detected by the X-ray imaging 
by Mushotzky et al.\ (2000), confirming the presence of a luminous AGN
in the source.

\subsubsection{SMMJ163639.01+405635.9}

This source is a good example of a heavily extincted starburst in an SMG
and was recently discussed by Swinbank et al.\ (2006).  This $z=1.485$
galaxy has weak \hb\ emission line with \hal/\hb\,$=10.4^{+29.6}_{-4.9}$. 
The \hal\ and \OIIIF\ emission lines are spatially extended 
($\sim1.2''$ or 10 kpc)(Figure~2). 
There is  only an upper limit on its X-ray emission, f$_{\rm 2-8KeV}<
2.2\times10^{-15}$\,erg\,sec$^{-1}$\,cm$^{-2}$ from Manners et al.\
(2003),
which does not strongly constrain the presence of a luminous AGN given
the
possibility of substantial absorption (e.g.\ Alexander et al.\ 2005). 
The possible detection of \OI\ emission line may hint at the 
presence of an AGN,  although the line ratios of \OI/\hal\,$\sim0.1$ and 
\OIIIF/\hb\,$\sim3.5$ can be explained by a relatively highly ionized 
starburst 
nebulae (Osterbrock et al.\ 1989).

\subsubsection{MMJ163655+405910}

This heavily obscured AGN at $z=2.605$) was found in the MAMBO survey of
Greve et al.\ (2004) (and is also called N2\,1200.18) and was detected in
X-ray imaging with {\it Chandra} (Manners et al.\ 2003).  It has broad,
FWHM$_{rest} \sim 2000$--2500\,km\,sec$^{-1}$, emission lines of \lya,
\CIV\ and \hal\ in the rest-frame UV and optical wavelengths, with a
high \OI/\hal\ ratio ($\sim 0.3$) (Willott et al.\ 2003; Swinbank et
al.\ 2006) which is typical of AGN (Osterbrock 1989).  Our data also
show asymmetric \hb\ and \OIIIE\ emission line profiles, which exhibit
``blue wings'' in their profiles. Such profiles have been interpreted
as evidence for wind activity from the AGN, although contribution from
other components is possible (Swinbank et al.\ 2006).

\subsubsection{SMM\,J221737.39+001025.1}
 
Our ISAAC spectrum shows strong, narrow \hal\, \hb\, \OIIID\ and \NII\ emission
lines at a redshift of $z=2.610$ (FWHM$_{\rm rest}$ of \hb\ is
$290\pm50$\,\kmsec).  To investigate the restframe optical properties,
we retrieved an archival $i'$-band image taken with Subaru Telescope's
Prime Focus Camera (Suprime-Cam) using
SMOKA\footnote{http://smoka.nao.ac.jp}.  The image shows an elongated
structure, $\sim1.3''$, towards the North-West and the spectrum was
taken with the slit aligned along the major axis of this source.  
We identify two
separate \hal\ emission lines with a velocity offset of $\sim
300$\,\kmsec and a spatial offset $\sim0.2$--$0.3''$ ($\sim 2$\,kpc).
These suggest the system is a merger.  The \hal\ and \hb\ emission
lines do not show asymmetric profiles or detectable broad line
components.

\subsection{Composite spectra} 

Since many of our individual spectra have modest signal-to-noise, we
have also constructed several composite spectra to
investigate the general properties of subsets of the SMG population.

We create the composite spectra by deredshifting each spectrum based on
redshifts measured from the \OIIIF\ lines, subtracting continuum
emission using a first order spline fit and averaging all of the
spectra with 3-$\sigma$ clipping after normalizing by \OIIIF flux.
We smoothed the higher resolution spectra taken at Keck and VLT to
match the low resolution Subaru spectra before stacking.  
Either stacking the spectra with weights based on their individual
signal-to-noise ratio or an unweighted stack does not alter any of the
conclusions below.  We derive a composite spectrum for 
those sources  which show QSO signatures
(``QSO''; i.e., classified as QSO) and for 
those galaxies
that individually show signs of an AGN in their optical spectra
(``OPT-AGN''; i.e., those classified as AGN in the column of ``Class''
under ``OPT'' category in Table~2).   The former is made from only three
individual spectra, while the latter comes from nine spectra.  The
resulting composite spectra are shown in Figure~3.  We do not make a
composite of starburst (``SB'') sources since there are only
two sources in our sample 
classified as ``SB'' or intermediate (``int'') from their
restframe optical spectra.
The details of the classification will be discussed in \S4.1. 

The emission lines of \hb\ and \OIIID\ lines are clearly seen in both
the composite spectra.    In
addition in  the ``QSO'' spectrum, many strong lines are visible,
including [Ne{\sc iii}]$\lambda3869$ and several Fe{\sc ii} lines at
$\lambda=4570$, 5167 and 5200--5360\AA, although the \OII\ line 
is only marginally detected.  By fitting a Gaussian to the
\hb\ and \OIIID\ emission lines, we measure the FWHM$_{\rm rest}$ of \hb\
as $\sim3200\pm1000$\,\kmsec\ after correction
for the instrumental resolution.  This is $\sim2000$\,\kmsec\ lower than the
average FWHM of QSOs at $z=0.1$--2.1 (Jarvis \& McLure 2006). The
\OIIIF/\hb\ ratio is $0.36^{+0.33}_{-0.18}$. All these spectral features 
are typical of type 1 AGNs studied locally.  

On the other hand in the composite
``OPT-AGN'' spectrum, a Gaussian profile fit to 
the \hb\ emission line yields FWHM$_{\rm rest}$ of
$1730\pm500$\,\kmsec\ 
(it should be noted that the \hb\ line fit is not improved 
by including a narrow line component due to the low spectral 
resolution of our spectra)
and \OIIIF/\hb ratio of
$3.2^{+1.0}_{-0.6}$, in addition the \OII\ line is well-detected.  
The \hb\ line, which is broader than typical type 2 AGNs, and relatively  
low \OIIIF/\hb\ line ratio, is similar to that of local NLS1 
(although by definition these should have \OIIIF/\hb/$<$3.0). The Fe{\sc ii}
emission lines, which are one of the characteristic 
features seen in local NLS1's,
are marginally detected with $\sim 2\sigma$ features seen around
5200 \AA\ in the spectrum, and we can see some marginal detections
in individual spectra (SMM\,J123549.44+621536.8, SMM\,J123635.59+621424.1,
SMM\,J163650.43+405734.5, and SMM\,J163706.51+405313.8), all of which
have 
broad
\hb\ emission of FWHM$_{rest}\sim2000$\,\kmsec (Figure~1).  The
resultant spectrum is consistent with a scenario where the restframe optical
spectra classified as ``AGN'' in the
UV in reality comprise two types: one has relatively
broad, $\sim 2000$\,\kmsec, FWHM for the \hb\ lines and the other has narrow
\hb\ lines with a relatively high \OIIIF/\hb\ ratio, typical of Type~2
AGNs.  
There are clearly differences in the extinction of the circumnuclear
region of these two types of objects implied by the difference in luminosity 
and spectroscopic properties of the restframe-UV emission, although
there is no systematic 
difference in the \hal/\hb\ ratio we measure for them.

\section{Discussions}

\subsection{Emission Line Diagnostics}

In Figure~4, we plot the observed \OIIIF/\hb\ versus \NII/\hal\
emission line ratios of the 13 galaxies in our sample for which we
have secure \hal\ detections and some information about \NII/\hal.
This diagnostic plot, termed the BPT diagram, can be used to identify
the source of gas excitation (Baldwin et al.\ 1981).  Based on this
diagram we classify the spectra into three types, starburst (SB),
intermediate (int) or non-thermal (AGN), as listed in
Table~2. We use the definitions from Kauffman et al.\ (2003) which are
derived for a large sample of local SDSS galaxies.  We classify the
sources between the boundary of Kauffman et al.\ and the classical
definition of Veilluex \& Osterbrock (1987) as ``int''.  We also classify
galaxies as AGN which have \hal\ and/or \hb\ FWHM$_{\rm rest}$ greater than
1500\,\kmsec, as it is difficult to understand the formation of such
large line widths
from gas motions in star-forming regions.  
This limit is also greater than the coarse spectral resolution
of OHS ($\sim1400$\,\kmsec).  For comparison we also
plot the emission line flux ratios
from local ULIRGs (Veilleux, Kim \& Sanders 1999)
and note that the SMGs in our sample occupy the same region of the
diagnostic diagram as local ULIRGs.  
The curves show various criteria for separating AGNs and
the star-forming galaxies (see Figure~4).  
It is clear that the majority, 8/13, of sources 
in our sample (including all but
one, SMM\,J163639.01+405635.9, with all 
four emission lines detected) 
are classified as AGN based on these criteria.  We reiterate that
this subsample may be biased towards strong line emitters 
(due to the requirement to have detected lines in our low-resolution
spectra) and
so this is perhaps not a surprising result. 

Moreover, we must interpret the BPT diagram with caution since
``superwind'' ejecta (shock-driven line emitting gas) can
occupy a very similar region to AGN (Dopita \& Sutherland 1995).  
To illustrate this possibility in more detail, we
plot the emission line ratio of the wind structure in M\,82 from
a $\sim1$\,kpc region (Shopbell \& Bland-Hawthorn 1998) and in
a $1.9\times4.3$\,kpc$^2$ area of NGC\,6240 (Schmitt et al.\ 1996).  The former
is indicative of a wind which is dominated by photoionization, and
the latter illustrates
the line ratios expected from shocks in a very dense environment.  
Some SMGs show
very similar emission line ratios to NGC\,6240, although most of them have
lower \NII/\hal\ and higher \OIIIF/\hb\ ratios. 
Further support for the wind scenario is that 
P-Cygni features are seen in the rest-UV emission lines (Chapman et
al.\ 2003a, 2005) of a significant fraction of the SMG population, 
supporting the presence of ``winds'' arising from the vigorous
starburst activity.  Indeed, the majority  (6/8) of this subsample (with four
detected emission lines) are classified as ``SB'' or ``int'' from their rest UV
spectroscopic features (Chapman et al.\ 2005; Table~2). 
 The power sources in
SMGs is discussed further in \S4.2.

Looking at the individual
sources in Figure~4, we note that SMM\,J123622.65+621629.7 has a very low \NII/\hal\
($<0.05$) emission line ratio and no detection of \OIIID\ and \hb\
(see also Figure~5 in Smail et al.\ 2004) with \hal/\hb $>$\,4.25 and
\OIIIF/\hal\,$< 0.24$.  This sources is an interacting system of a
relatively blue($B-R\sim 0.2$) galaxy with extremely red ($I-K=4.0$)
companion, where the latter is a hard X-ray source.  
Whilst the slit was aligned
along the major axis of the red X-ray source, it is possible that it 
also passed through the blue component and the line emission may
be contaminated. To avoid biasing our sample, we have
therefore eliminated this source
from our subsequent analysis and discussion.

\subsection{Extinction and Hidden Star Formation} 

SMGs are dusty systems with large dust masses, $>10^8$\,M$_\odot$, 
and high bolometric luminosities
($>10^{12}\lsun$). The
presence of large quantities of dust and its associated
reddening may also explain the large
discrepancies between the SFRs derived for SMGs 
from their far-infrared and \hal\
luminosities (Swinbank et al.\ 2004), which imply extinction in \hal\ of
factors $\sim10$--100.
There are of course alternative explanations: that
the bulk of the far-infrared emission originates from other sources which are 
too dusty to see even at restframe optical wavelengths, such as 
very highly obscured AGN, or due to 
emission which falls outside of the slits used in the \hal\ measurements. 
Although, the latter explanation is unlikely
as these observations are based on radio-identified 
sources with precise positions ($\sim 0.5''$; Chapman
et al.\ 2005), and so it is unlikely that
a major source of bolometric emission has been missed
by the observations.  

To investigate the internal reddening of SMGs
(at least for those regions which are visible in the restframe optical) 
we plot the \hal/\hb\ ratios  as a function of their
far-infrared luminosities in Figure~5. To calculate $A_V$, 
we use the reddening curve from 
Calzetti et al.\ (2000), and assume an intrinsic \hal/\hb\ 
ratio of 3.0, which is between the values for
typical Seyfert 2 galaxies and/or LINERs (3.1, Halpern \&
Steiner 1983; Gaskel \& Ferland 1984) and star-forming galaxies,
2.85 (Veilleux \& Osterbrock 1987). The 
observed \hal/\hb\ ratio for the SMGs is typically 5--20 and the derived
extinction spans $A_V=1$--4 with a median value 
of $2.9\pm0.5$ (where the
error comes from bootstrap resampling). This estimate is
consistent with the results based on the spectral energy distribution (SED) 
fitting of optical to near-infrared photometric data (Smail et al.\ 2004), and
slightly higher than that 
derived from optical to mid-infrared SEDs ($1.7\pm0.3$, Borys et al.\
2005) where the latter did not include any contribution from Thermally
Pulsed-AGB stars in
the model SEDs, which might lead to an underestimation of the
reddening (Maraston 206). 

In Figure~6 we compare the extinction corrected SFRs derived from the
\hal\ and far-infrared luminosities.  
The far-infrared luminosities come from Chapman et al.\ (2003b; 2004) 
based on SED model fitting to the observed 850$\mu$m and 1.4-GHz fluxes 
at their known redshifts, assuming the local far-infrared-radio correlation 
holds (Condon et al.\ 1991; Garrett 2002).  
We also include observations
for local {\it IRAS} galaxies (Kewley et al.\ 2002) and {\it ISO}
galaxies (Flores et al.\ 2004).  The typical $A_V$ in these samples are
$\sim0.5$ and $\sim2.4$, respectively.  The extinction corrected \hal\ 
luminosities for the {\it IRAS} and the {\it ISO} galaxies are all 
calibrated in the same manner as for our SMG samples based on their 
\hal/\hb\ ratio.  The SFRs from the \hal\ and from the far-infrared 
luminosities are derived using
the equations given in Kennicutt (1998). 

The  correlation between
the far-infrared and reddening-corrected \hal\ luminosities
(Figure~6)  appears to be relatively good with a 
linear relation extending over five orders of magnitude in SFR,
although with some scatter, with the
most
luminous SMGs in our sample having SFRs 
approximately an order of magnitude higher than
those of the brightest {\it ISO} galaxies.  
The good agreement between
the two SFRs when using the reddening-corrected \hal-estimate
confirms that the discrepancies between the SFRs 
seen in Swinbank et al.\ (2004) 
are in large part
due to dust extinction and moreover that
the bulk of the far-infrared luminosity 
in these galaxies is probably derived from star formation.
We note that it is likely that slit-losses and placement contribute
to the scatter in these measurements as we are combining observations
of \hb\ and \hal\ from different telescopes and instruments.
For example, our brightest far-infrared source,
SMM\,J163650.43+405734.5 (N2\,850.4), has a lower SFR measured from
\hal\ than from the far-infrared.  However, this galaxy is spatially extended
and has a very complex structure in the restframe optical (Smail
et al.\ 2003; Swinbank et al.\ 2005).  It is therefore likely  
that our slit covered only a part of the \hal\ emitting region.
Finally, any remaining systematic offset between the two SFR estimates
may  be caused by the fact that our A$_V$ estimates
only reflect the reddening to  the 
optically detectable gas and thus are
not necessarily a good indicator 
of the total column towards the bolometric sources in these objects.

\subsection{Growing Black Holes in the SMGs}

\subsubsection{Spectral Similarity with Narrow Line Seyfert 1 Galaxies}

SMGs are proposed to be the progenitors of present-day massive
spheroidal galaxies, because of their high star-formation rates and 
their large stellar, gas and dynamical
masses (Smail et al.\ 2004; Neri et al.\ 2003; 
Greve et al.\ 2005; Borys et al.\ 2005; Tacconi et al.\ 2006;
Swinbank et al.\ 2006). 
Most massive galaxies in the local Universe contain
super-massive black holes (SMBHs) (e.g., Ferrarese \& Merritt
2000; Gebhardt et al.\ 2000; Marconi \& Hunt 2003; Heckman et al.\
2004). Equally an AGN appears to be almost universally present in
SMGs: based on the extremely
sensitive X-ray observations of the {\it Chandra}
Deep Field North (CDFN), Alexander et al.\ (2003; 2005a,b) found more
than $\sim 75$\% of SMGs to be detected in hard X-rays,  indicating
they contain an accreting SMBH.  It is therefore interesting to estimate 
the mass of, and accretion rates onto,
the central black holes of SMGs to constrain the coevolution of
the SMBHs and the  stellar masses of their surrounding bulge 
(Kawakatu et al.\ 2003;
Granato et al.\ 2004).  

The three sources classified as ``QSO'' in our sample have characteristics 
typical of local NLS1: low \OIIIF/\hb\ ratios ($\sim 0.5$--1.8), 
and detectable Fe{\sc ii} emission in 
their individual and also composite spectra (Figures~1 \& 3). NLS1 are commonly interpreted as hosting
rapidly growing SMBHs (Collin \& Kawaguti 2004), and hence the 
spectral similarities of 
these SMG-QSOs with local NLS1s could imply comparable physical conditions 
in the accretion disk around the SMBH in the SMGs. However,
the SMGs  have FWHM$_{\rm rest} \sim 2000$--2500\,\kmsec\
for their Balmer emission lines, and so
they are not formally NLS1s because 
these lines width are higher than the definition used for NLS1
(FWHM$_{\rm rest}$ of 
\hb\ of $< 2000$\,\kmsec).  Nevertheless, it is however worth noting that their 
\hb\ FWHM$_{\rm rest}$  are close to the minimum for QSOs at $z\sim 0.1$--2.1 
(Jarvis \& McLure 2006), and narrower than the average of radio 
quiet/loud QSOs($\sim 4800$--6500\,\kmsec). Although, we caution that
with the limited signal to noise in our spectra
we may underestimate the line widths, missing weak and broader line components.
For instance, in the composite spectrum of SMG-QSOs we estimate the
FWHM$_{\rm rest}$ of the \hb\ line as $3200\pm1000$\,\kmsec\ 
using a single Gaussian 
fit. This is 500--1000\,\kmsec\ broader than the mean of 
the individual spectra, suggesting there may be an undetected broad
component present in them.
More secure estimates of the 
line widths would either need
observations of stronger emission lines such as \hal\, which are not
available for these sources, or much deeper observations.

The FWHM$_{\rm rest}$ of the Balmer emission lines in those SMGs with AGN-like
features (but omitting the three
sources classified as ``QSO''), are 1000--3000 
\kmsec. They are at least 1000--2000 \kmsec\ lower than 
the average FWHM of QSOs at $z\sim 0.1$--2.1 
measured from \hb\ and/or Mg{\sc ii} lines
(Jarvis \& McLure 2006) suggesting that the SMGs host lower
mass SMBHs.  This would support the claims of Alexander et al,\
(2005a,b;
see also Borys et al.\ 2005) based on Eddington-limited assumptions.

The similarities of the rest-frame optical spectral features of some SMGs
to NLS1s implies rapid growth of the SMBH in SMG's nuclei. A total of 
5/9 of the SMGs classified as ``AGN'' in our sample have relatively narrow FWHM$_{rest}$ 
(up to $\sim 1600$--3700\,\kmsec) for their \hal\ or \hb\ emission lines, and 
3/5 show marginal Fe{\sc ii} emission. 
Therefore, the 
Eddington-limited accretion determined for local NLS1 galaxies may also be
appropriate for SMGs. Assuming this, the measured line-widths are
then consistent with the estimate of the central 
BH masses derived from their X-ray luminosities under the assumption of 
Eddington-limited accretion ($\sim 10^{6-8} \msun$, Alexander et al.\ 2005a).  
However, this conclusion appears to be undermined
by the fact  that three of these
NLS1-like SMGs display high ($>10$) \OIIIF/\hb\  
ratios which far exceed 
the NLS1 definition of \OIIIF/\hb\ $<3$, 
and thus these comparisons may not be appropriate. 

\subsubsection{The origin of \OIIIF\ excesses in SMGs}

To further test the claim that SMGs have small SMBH masses 
we compare the \OIIIF\ and hard X-ray 
luminosities. There is a well-studied correlation between the hard X-ray
and the optical \OIIIF\ emission line luminosities in local AGN
(e.g.\ Mulchaey et al.\ 1994).  This correlation can be used to gauge the
black hole masses and the accretion rates of AGNs within our sample.

In Figure~7, we show the hard X-ray versus \OIIIF\ luminosities of the
SMGs (uncorrected for any extinction/absorption).  All 22 SMGs in our sample
have  hard X-ray coverage, but of varying depth: CDFN: Alexander et al.\ (2003),
CFRS\,03hr: Waskett et al.\ (2004), SSA13: Mushotzky et al.\ (2000), SSA22:
Basu-Zych \& Scharf (2005), and ELAIS N2: Manners et al.\ (2003). We
adopt the hard X-ray fluxes from these observations, although 9/22 of them
yield only the upper limits.  For comparison, we also plot observations of
local ULIRGs (Ptak et al.\ 2003; Franceschini et al.\ 2003), as well as
Seyfert 1 and Seyfert 2 galaxies and the PG QSOs, representative 
of more luminous type 1 AGNs (Alonso-Herrero et al.\ 1997; 
Mulchaey et al.\ 1994).  All of these comparison samples are
the {\it observed} luminosities: there are no extinction corrections
applied to either the X-ray or \OIIIF\ measurements.

Figure~7 also shows the relation for Seyfert 2 galaxies suggested by
Mulchaey et al.\ (1994). Compared to the QSOs and Seyfert 1
galaxies, which are selected to represent unabsorbed hard X-ray
sources, the majority of our SMGs are typically an order of magnitude
brighter in \OIIIF\ for a given hard X-ray
luminosity.  We note
that a similar excess of \OIIIF\ emission is also seen in local
ULIRGs. 

Could this apparent excess be due to absorption/extinction?
The typical Hydrogen column densities to the AGN in SMGs have been
determined by Alexander et al.\ (2005b), yielding
N$_H \sim 10^{23-24}$\,cm$^{-2}$, 
with corrections to
their hard X-ray luminosities of 2.5--20\,$\times$.  
Equally, the typical \OIIIF\ luminosity
correction, adopting the extinction estimated from 
the Balmer decrement, $A_V \sim 2.9$, is also approximately a factor ten:
${\rm F}_{{[OIII]{\lambda5007}}_{corrected}}
= {\rm
  F}_{{[OIII]{\lambda5007}}_{obs}}\cdot((H\alpha/H\beta)/(H\alpha_{0}/H\beta_{0}))^{2.94},$ 
where H$\alpha_{0}$/H$\beta_{0}$ is assumed to be 3.0 (see Bassani et
al.\ 1999).   Unfortunately we
have only one source (SMM\,J123549.44+621536.8) with reliable
estimates of the H{\sc i} column density
and reddening correction  which has $10\times$
and $28.5\times$ corrections to the hard X-ray and \OIIIF\ luminosities
respectively and with yields corrected
luminosities of $1\times 10^{44}$ \ergsec\ and $6.2\times10^{44}$
\ergsec\ respectively (Figure~7).  

As the extinction corrections for \OIIIF\ and hard X-ray luminosities 
run parallel to the trend in Figure~7, the \OIIIF\ 
excess can not be explained by a simple reddening effect. 
We also caution that the reddening corrections applied
to the \OIIIF\ fluxes are uncertain since  \OIIIF\ may arise in
external shocks which suffer much less extinction than the
H$\alpha$/H$\beta$ ratio suggests.  
We also note that the apparent \OIIIF\ excess could arise  simply due to the
relatively shallow X-ray coverage in several of our fields where
the sources only have upper-limits on their hard X-ray fluxes. 
However the fact that 3/4 sources with \OIIIF\ and hard X-ray
detections from the CDFN, which has by-far the best X-ray data, 
show the excess provides good evidence for the reality of this feature. 

While some of the \OIIIF\ flux we see arises from the obscured AGN, we
suggest that the excess \OIIIF\ flux arises, at
least in part, from shock-induced (``superwind'') activity.
There are some cases of plausible ``superwind'' driven \OIIIF\ excesses
seen in our SMG sample as seen by the structured 
\OIIIF\ line profiles (asymmetric/broad/multi-peaked) 
and the spatially extended emission (Figures~1
\& 2; see also Smail et al.\ 2003).  

In order to examine the possibility that shock-induced gas causes the
excess \OIIIF\ emission, we first  search for the signature of shock-excited
nebular emission using the simple criterion of
\NII$\lambda6583$/\hal\,$\geq1$ and 
\SII$\lambda\lambda6716,6731$/\hal\,$\geq 0.5$. 
Using the line ratios from
the stacked spectrum of SMGs in Swinbank et al.\ (2004)
we find that they lie outside of this shock induced criterion.  However, this
criterion is only valid for the shocks with large outflow velocities
and a relatively weak starburst radiation field (Veilleux et al.\ 2005)
and therefore may not be applicable to the SMGs.  Another test is
to use the line ratios of \OIIID/\OIIIG\ and/or
\NII${\lambda\lambda6548,6583}$/\NII$\lambda5755$, which can be used to
estimate the temperature of the nebular gas. 
These ratios will provide robust estimates of electron temperature of 
the emission nebulae, yielding high ($\sim30,000$\,K) temperatures if 
the gas is ionized mainly by shocks (as in the Cygnus Loop) and
lower temperature ($\sim 15,000$\,K) for photoionization-dominated clouds 
seen in star-forming regions (Osterbrock 1989). As these methods rely on 
measurements of relatively weak emission lines  
\OIIIG\ and \NII$\lambda5755$, only our composite 
spectra have sufficient signal to noise to be useful. From the composite spectrum in 
Figure~3 and also from the total 
SMG composite spectrum 
in Swinbank et al.\ (2004), we derive an upper limit on 
\OIIID/\OIIIG\,$<37.2$, and 
\NII${\lambda\lambda6548,6583}$/\NII${\lambda5755} < 23.6$.
Both ratios imply an upper limit to 
the electron temperature of less than 20,000\,K.
This is consistent 
with  the expected temperature in photoionization-dominated clouds (with an electron density of 
$\sim10^{3-4}$\,cm$^{-2}$). If the electron density is higher than 
this, collisional de-excitation begins to play a role and 
the estimated temperature is reduced. 
These results would appear
to rule out the dominance of shock excitation similar
to that seen in galactic supernova 
remnants. 

As described in Dopita \& Sutherland (1995), the optical line ratios of
Seyfert 2s can also be explained by fast (300--500\,\kmsec) shocks, if
the precursor H{\sc ii} regions in front of the shock absorb
most of the UV photons generated by the shocks.  The calculated
electron temperature  is $\sim17,000$\,K for this
``shock + precursor'' model from Dopita \& Sutherland (1995), which is
consistent with the limit on the
electron temperatures in SMGs estimated from our \NII\ and
\OIIIA\ emission line ratios.  Thus there is a plausible origin for the
\OIIIF\ excess we see compared to typical AGN: shocks associated
with supernova explosions in relatively dense gas environments, 
where the precursor H{\sc ii} clouds are still present.
Thus we suggests that those sources with high ($\sim5$--10) \OIIIF/\hb\ ratios 
and broad ($\sim$2000 \kmsec) FWHM of \hb\ lines can be explained by 
a combination of a NLS1-type AGN residing 
in an environment of shocks associated
with supernova explosions in relatively dense gas. This would
explain all their observable properties, including the  
high \OIIIF/\hb\ ratios (Dopita \& Sutherland 1995). 

\section{Conclusions}

Using near-infrared spectroscopy we have observed the redshifted \hb\,
the \OIIID\ and \OII\ emission lines in a sample of 22 Ultra-luminous 
Infrared Galaxies at high redshifts. Twenty of the
sources in our sample are submillimeter galaxies at $z\sim 1.0$--3.5.  
Combining our observations with previous studies of
the \hal\ and the \NII\ emission 
from these galaxies and also with observations of their
hard X-ray and far-infrared emission, 
we have placed constraints on the physical properties of this population. 
We conclude the following:

\begin{enumerate}

\item A majority 
   of our sample (14/22) have spectra which are classified as ``AGN'' or ``QSO'' 
  based on several restframe optical spectroscopic diagnostics. Specifically, 
  for those sources with detections of the
 four emission lines necessary to construct a 
  BPT diagram, 8/9 are classified as ``AGN''. It should be noted that 
  there is no confirmed pure starburst galaxy in our
  sample, although several sources show intermediate spectral properties.  
  This is likely to be  caused by our sample selection, which is biased towards 
  galaxies with bright near-infrared magnitudes and also 
  to those exhibiting
strong line emission. Thus we caution that our results should not be
  taken as representative of the whole SMG population. 

\item Using the \hal/\hb\ flux ratio we are able to
  estimate the internal extinction in our SMGs.  We measure a median
  extinction of $A_V=2.9\pm0.5$, which is similar to the extinction
  measured in local ULIRGs.  This value is also consistent with the
  estimates from the SED fitting in the restframe UV/optical 
  which are derived under the assumption of a dominant
dust-reddened young starburst (Smail et al.\ 2004).

\item We compare the SFRs derived from the dust-extinction-corrected
  \hal\ luminosities with those derived from the far-infrared luminosities, and
  find reasonable consistency between these for most of the SMGs in
  our sample.  The
  fact that the corrected \hal-derived SFRs correspond closely to those
 estimated from the far-infrared
   suggests that star-formation is the major contributor to the far-infrared
  luminosities in SMGs.

\item At least 11/19 of the SMGs in our
sample show a clear excess in the
ratio of their \OIIIF\ to X-ray luminosities 
  relative to values  for local AGNs. The five sources with the 
  highest \OIIIF/\hb\ ratios ($>10$), which are classified as ``AGN'' from our spectral 
  diagnostics, show this \OIIIF\ excess. 
One possible explanation for the 
  \OIIIF\ excess is that it is produced by ``Compton-Thick'' AGNs.  
However, this is
  inconsistent with the column density measurements (N$_H$) from
  fitting of the X-ray spectra for the sources in CDFN  and
  we argue that this is unlikely in most SMGs.
Instead, we suggest that
  the most plausible cause of
  the \OIIIF\ excess is shock-induced emission 
  arising from vigorous star formation (``super-wind'' activity).  
This scenario is supported in
  several galaxies by spatially extended and/or distorted/multiple
  \OIIIF\ emission line profiles.  Furthermore, using limits
 on the electron temperatures  
  from \OIIIA\ and \NII\ emission line ratios, we can explain the
 excess \OIIIF\ emission as arising from shocks in dense regions
  within these systems.  

\item The Balmer line widths in 9/22 sample galaxies exhibit broad
  emission components with relatively small FWHMs
 ($\sim1500$--3700\,km\,sec$^{-1}$). Three of them are classified as
  ``QSO'', but have smaller \hb\ FWHM (2100--2600\,\kmsec) 
  than are typical for QSOs. They also have lower \OIIIF/\hb\ ratios 
  and relatively  strong 
  Fe{\sc ii} emission, both of which are characteristics of local Narrow
  Line Seyfert 1s. 
  Among the other six sources, only one shows a low  \OIIIF/\hb\ ratio, 
  and four show high \OIIIF/\hb\ ratios (larger than seen in NLS1's). 
However, the high \OIIIF/\hb\ ratios may arise 
  from \OIIIF\ excesses due to shock excitation and hence
  removing this contribution would yield lower ratios 
  more consistent with NLS1 classification. Several of these sources 
  also have tentative evidence for Fe{\sc ii} emission, again
  characteristic of NLS1s. Thus, once account is taken of the
 potential contribution from shocks to the excess \OIIIF\ emission, 
  there appears to be close similarities between SMGs and NLS1s.
  The spectral classification of SMGs as NLS1s 
  may then indicate (as has been claimed for local NLS1s) 
  that SMGs have small mass black holes which are 
  rapidly growing at high accretion rates (Alexander et al.\ 2005ab;
  Borys et al.\ 2005). 
  Deeper spectroscopic observations 
   are essential to search for any obscured broad Balmer
  lines which might indicate larger SMBH masses and confirm the
  presence of  Fe{\sc ii} 
  lines which are common in the NLS1s. 

\end{enumerate}

Summarising our results: we conclude that  our sample of
SMGs  contains a population of vigorously star-forming galaxies
with high SFRs and strong extinction.  The activity in these
systems is driving shocks through the dense gas reservoirs
they contain and some of this material is being expelled
from the galaxies.  In addition, many of our sources show
evidence for low-mass, but rapidly growing, super-massive
black holes.  These results confirm the critical place of
the submillimeter-bright phase in defining the properties
of massive galaxies forming at high redshifts.

\acknowledgments 

We are grateful to Michael Balogh, Bob Nichol, Chris 
Miller and Dave Alexander for their providing invaluable information
and discussions.  TT and KS are also thank to all staffs of Subaru
telescope, especially to Dr. Kentaro Aoki and Dr. Takuya Fujiyoshi for
the supports on our Subaru/OHS observation.  
We also thank to the anonymous referee for the various comments and 
suggestions to improve our manuscript. 
IRS acknowledges support from the Royal Society. JEG acknowledges support
from a PPARC postgraduate studentship. AMS acknowledges a PPARC
fellowship.

\clearpage

%
%
\begin{figure}
\includegraphics[scale=0.9]{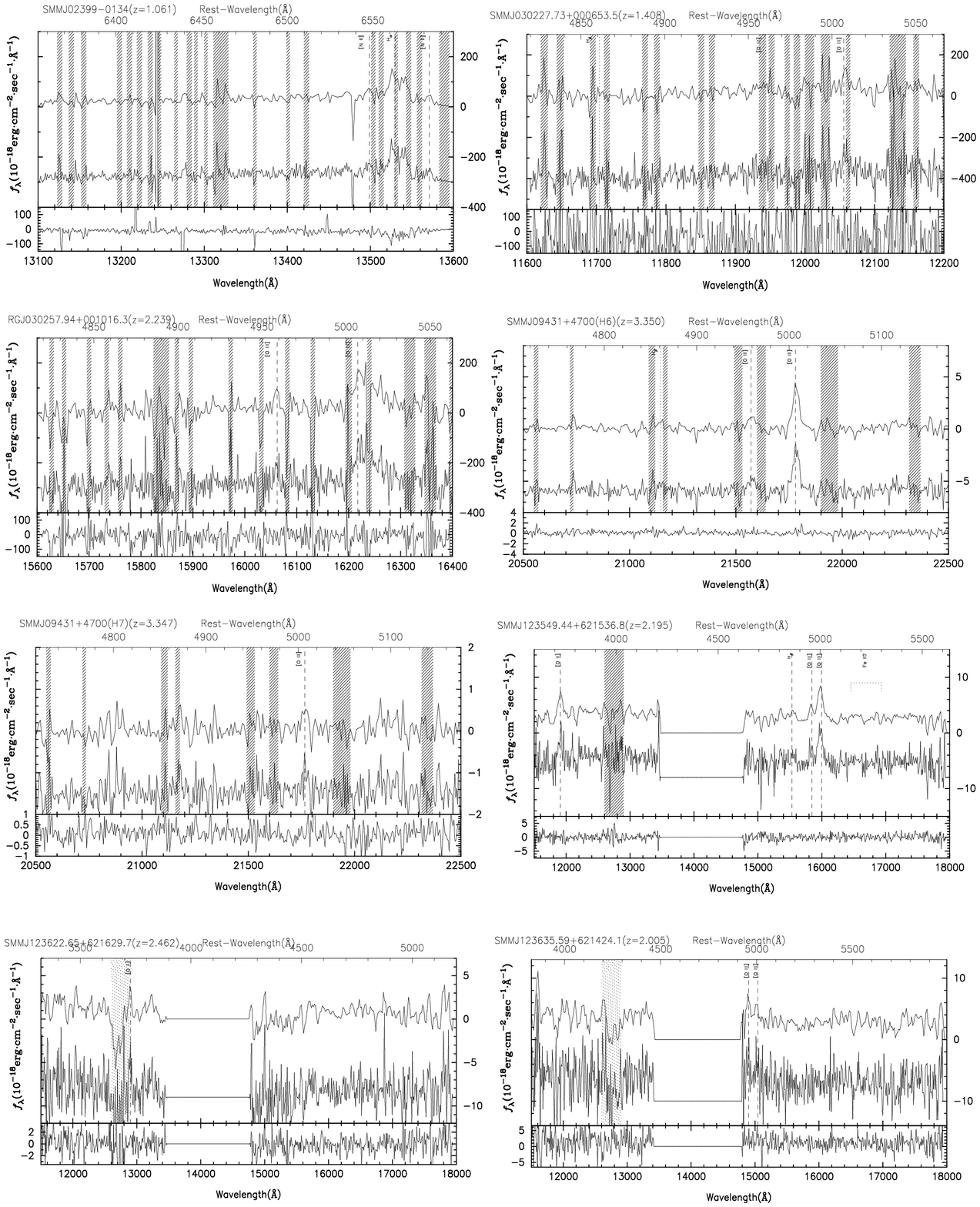}
\caption{The near-infrared
spectra of the galaxies in our sample. The upper spectrum in
  each panel is smoothed to the instrumental resolution, the middle
  spectrum shows the raw data and  the lower
  spectrum illustrates the sky emission as a function of wavelength. 
  The dashed and dotted lines show the detected emission lines 
  with $>4\sigma$ and 2--$4\sigma$ significance respectively. 
The upper axis gives the restframe wavelength scale at the source
redshift.  The shaded regions are areas effected by strong sky emission
or absorption.    
  \label{fig1}}
\end{figure}
\begin{figure}
\includegraphics[scale=0.9]{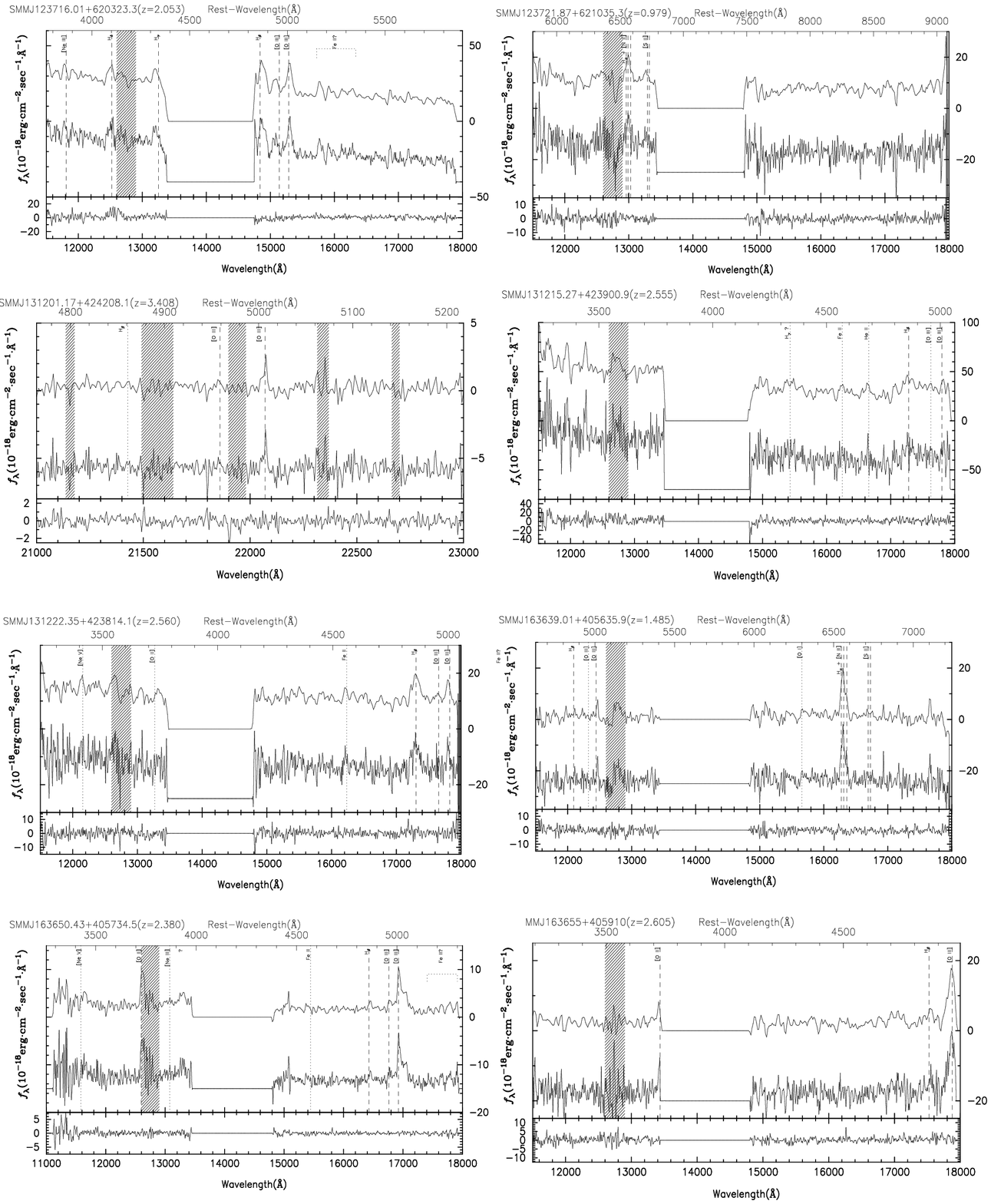}
\end{figure}
\begin{figure}
\includegraphics[scale=0.9]{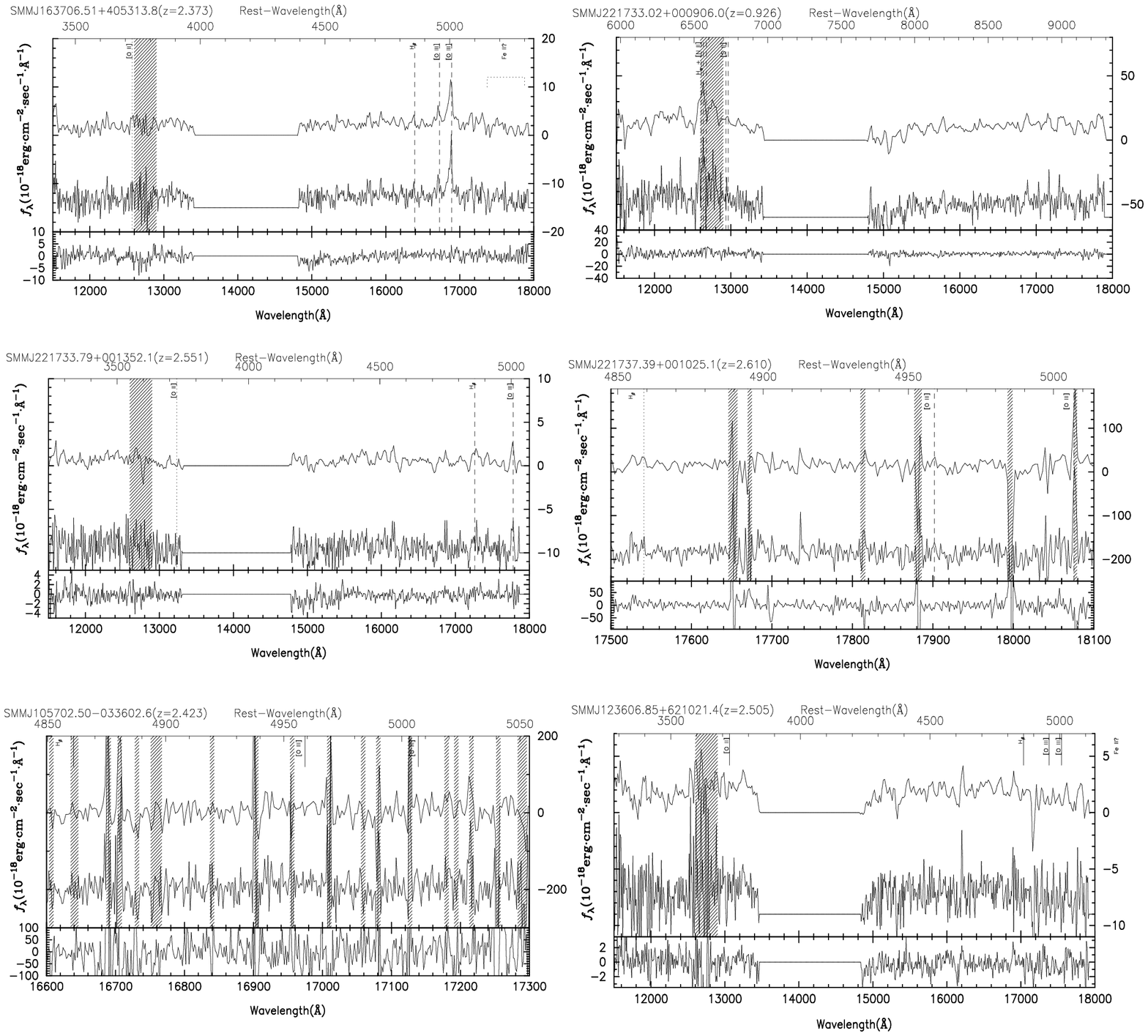}
\end{figure}
\clearpage 

%
%
\begin{figure}
\plotone{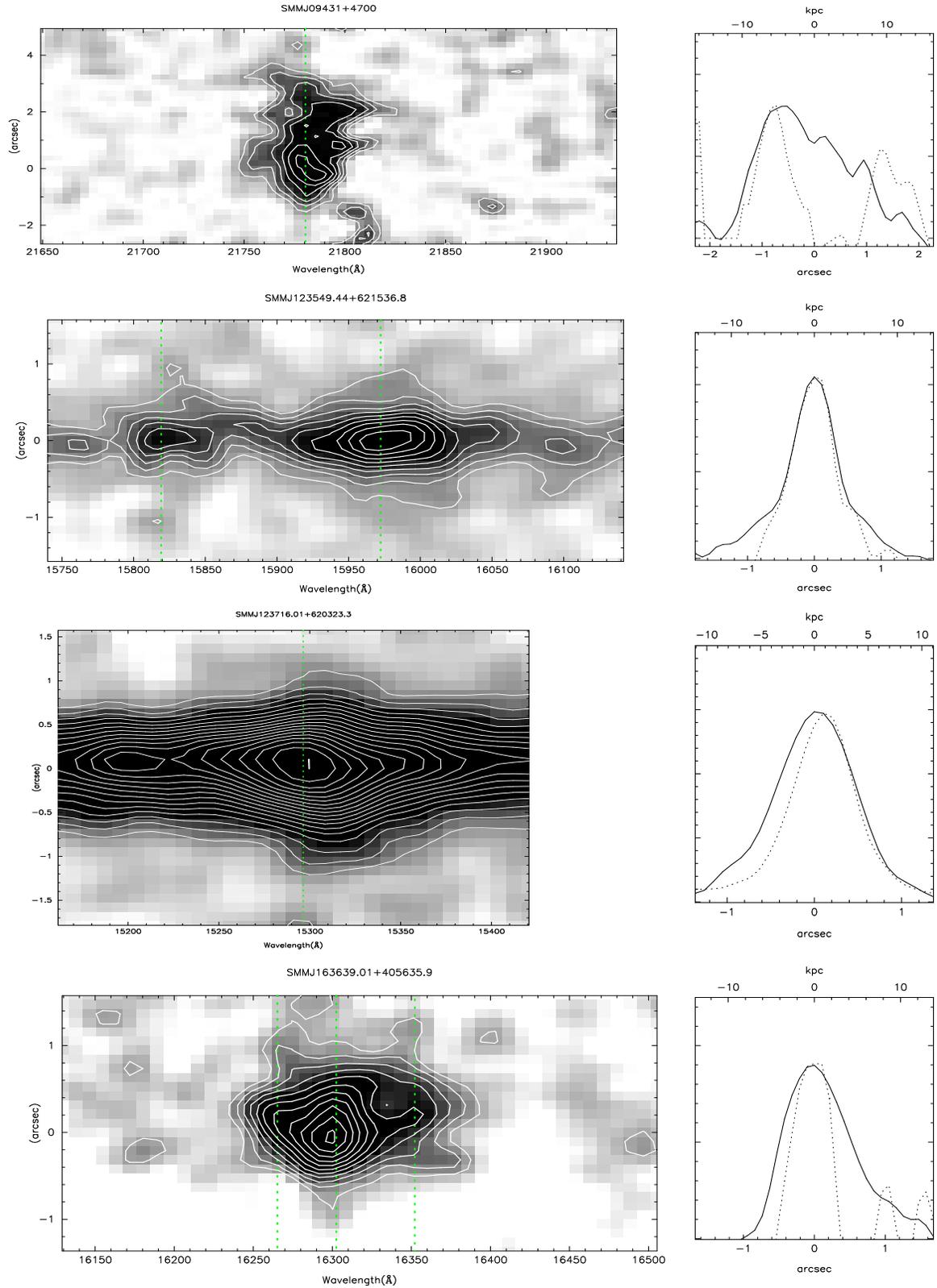}
\caption{
Position--wavelength maps (left) and slit profiles (right) for sources with 
spatially extended \OIIIF\ or \hal\ lines.  
~From top to bottom, these are SMM\,J09431+4700 (H6), SMM\,J123549.44+621536.8, 
SMM\,J123716.01+620323.3, and SMM\,J163639.01+405635.9. 
In the position--wavelength maps, 
the contours are spaced from 2$\sigma$ at 1-$\sigma$ intervals. 
The dotted lines display the expected wavelengths of 
the \OIIID, \hal\ or \NII\ 
lines at the redshifts listed in Table~2. 
In the slit profiles, the solid and dotted lines show the
light profile for the emission lines 
and neighboring continuum respectively.
The emission line profiles consist of \OIIIF\ emission for SMM\,J09431+4700 (H6), 
SMM\,J123549.44+621536.8 and SMM\,J123716.01+620323.3, and \hal\ emission 
for SMM\,J163639.01+405635.9. The width used for constructing the emission profiles 
correspond to 1000 \kmsec in their restframe. 
The spatially extended emission in \OIIIF\ and \hal, relative
to the neighbouring continuum, 
is likely to represent outflows of gas from these systems on
scales of $\sim 10$\,kpc.
}
\end{figure}
\clearpage

%
%
\begin{figure}
\plotone{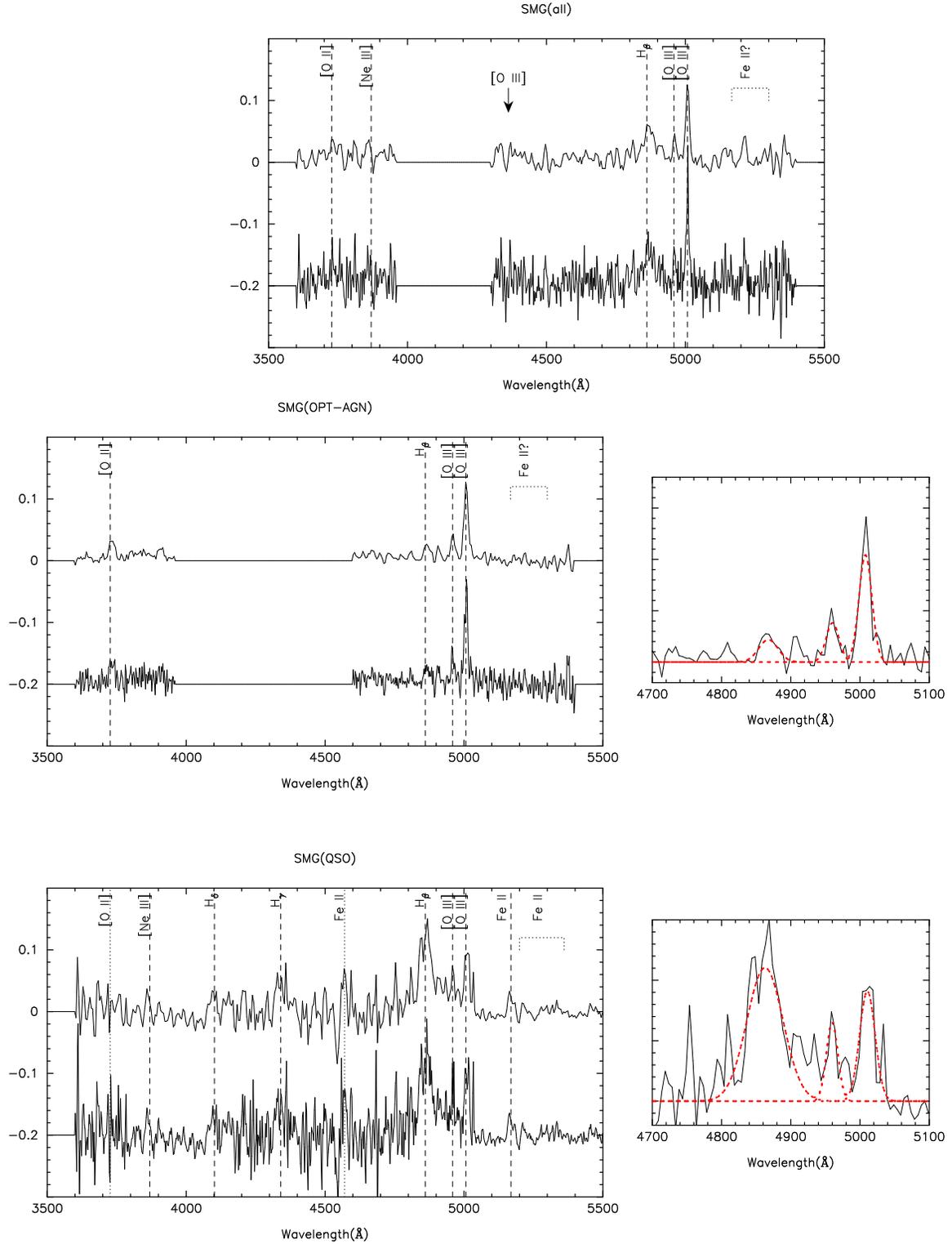}
\caption{The stacked restframe spectra of of subsets of galaxies in our sample.  
The top panel shows the
  spectrum of all the sources in our sample, 
the middle panel shows the combined spectrum of those sources
classified as AGN based on their restframe optical spectra (except for those 
classified as QSO). 
The bottom panel is the combined spectrum    of the three QSOs in our
sample. 
The upper spectrum in each panel is smoothed to the instrumental resolution 
of Subaru/OHS, and the lower one show the raw stacked spectrum. 
The right panels on the middle and bottom rows show 
  the region around the \hb\ and \OIIID\ emission lines with the best fit to these 
  emission lines using three Gaussian profiles overlaid (dotted lines). The
relative variation in the strength of \OIIID\ and \hb\ is clearly
visible between the different subsets. }
\end{figure}
\clearpage 

%
%
\begin{figure}
\plotone{f4.eps}
\caption{The BPT classification
diagram (Baldwin et al.\ 1981) for the galaxies in our
  sample. We plot the emission line flux ratio \OIIIF/\hb
  versus \NII/\hal for galaxies with strong \hal\ emission.  In cases
  where there is no detection of either \hb\ and \OIIIF\, we set
  \OIIIF\/\hb\ as 1.0 and highlight the
sources by plotting both lower and upper limits. Small open
  circles represent local ULIRGs from Veilleux et al.\
  (1999).  The dotted curve show the division between AGN and
  star-forming galaxies, and dotted line is for separating Seyferts and
  LINERs, as defined by Veilluex \& Osterbrock (1987).  Dashed and
  dash-dotted curves are the division between AGN and star-forming galaxies
  defined by Kauffmann et al.\  (2003) and Kewley et al.\ (2001),
  respectively.  We also plot  observations of 
 three SMGs from the literature 
  (Frayer et al.\ 2003; Simpson et al.\ 2004; Motohara et al.\ 2005). 
  Large open circles denote X-ray detected SMGs.
  For comparison we also plot the line ratios for NGC\,6240 (Schmitt et
  al.\  1996) and M82 (Shopbell et al.\ 1998) representing galaxies
  with (super)wind activity. Many of the galaxies in our sample 
show spectral line characteristics typical of AGN. 
However, the reader should note that our sample is not representative
of all SMGs due to the observational biases in our survey.
}
\end{figure}
\clearpage 

%
%
\begin{figure}
\plotone{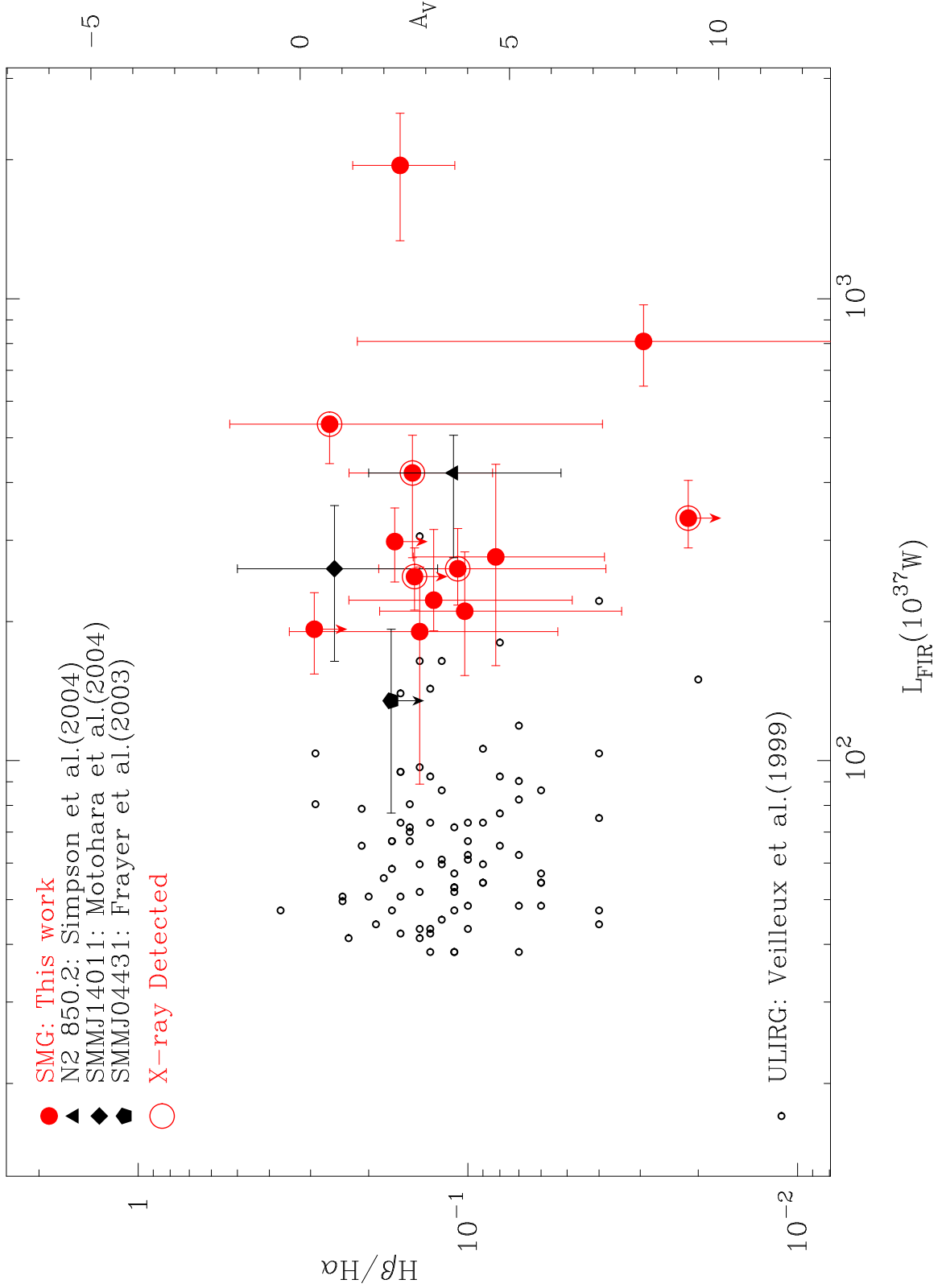}
\caption{
The \hb/\hal ratio as a function of 
far-infrared luminosity in our sample. 
The high-redshift ULIRGs display a 
similar  \hb/\hal\ ratio and range as typically-lower-luminosity
systems at lower redshift.  This suggests that they have comparable
levels of dust obscuration to their optically-detectable emission
line gas.
The symbols are
the same as in Figure 4.}
\end{figure}
\clearpage 

%
%
\begin{figure}
\plotone{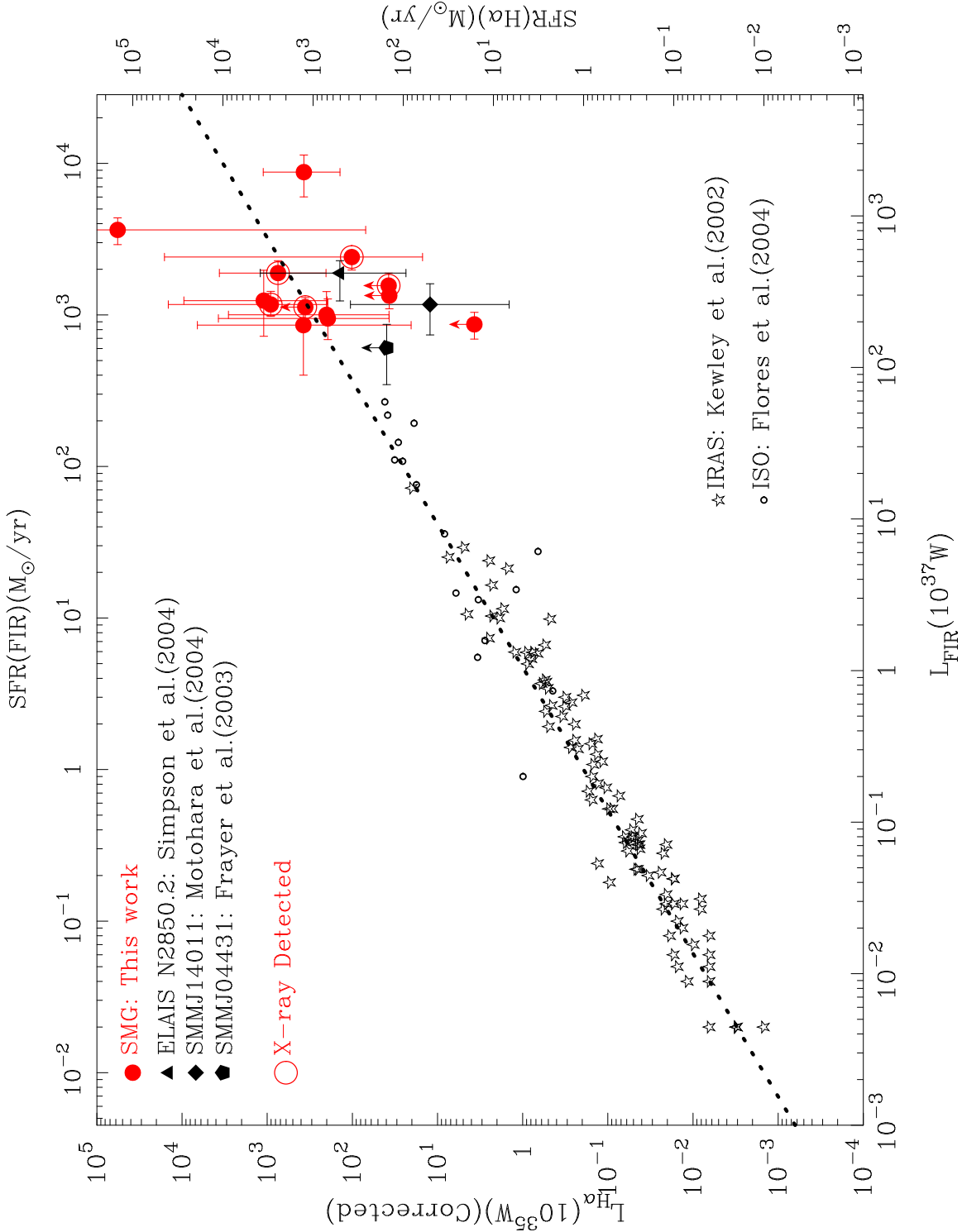}
\caption{The 
SFRs for our sample
derived from the extinction-corrected \hal\ luminosities
versus those estimated from the 
far-infrared luminosities.
For comparison we also plot 
samples of local {\it IRAS} galaxies (Kewley et al.\ 2002) and 
$z\sim 1$ {\it ISO} galaxies (Flores et al.\ 2004) 
as stars and open circles. 
The dotted line shows equality in the  Far-infrared and \hal\
SFRs. The high redshift sources appear to extend the agreement
between these two star-formation indicators to higher SFRs,
although there is considerable scatter (in part due to our 
use of \hal\ and \hb\ observations taken with different
instruments at different times).  Nevertheless, the broad
agreement between the \hal\ and far-infrared SFRs suggests
tha the bulk of the far-infrared luminosity in these galaxies
is derived from star formation.
}
\end{figure}
\clearpage 

%
%
\begin{figure}
\plotone{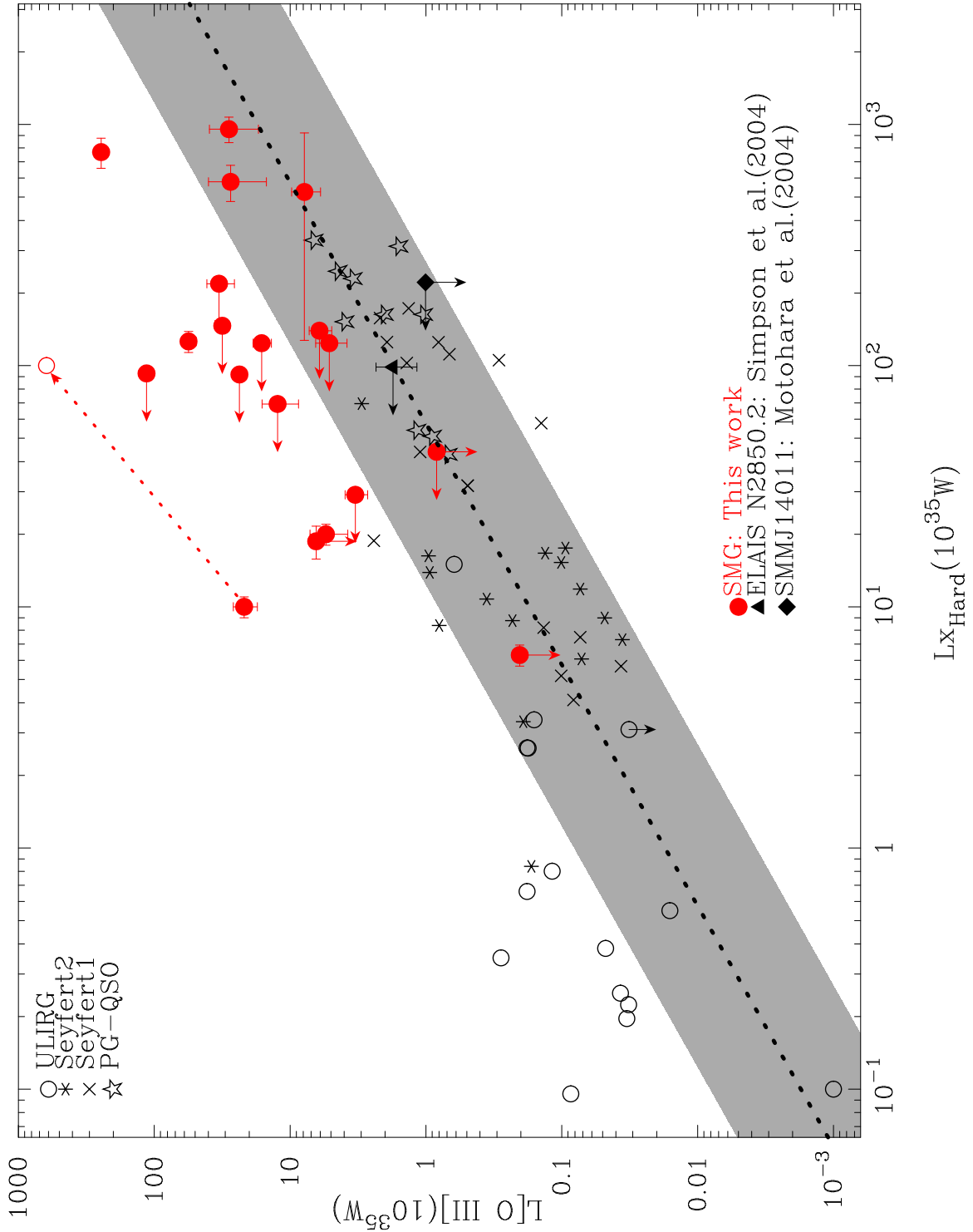}
\caption{Hard X-ray luminosities versus \OIIIF\ luminosities
for the galaxies in our sample. We also
  show similar observations for
local ULIRGs (triangles), Seyfert 2s (asterisks),
  Seyfert 1s (crosses) and local QSOs (stars) for comparison.  The dotted line
  shows the correlation between hard X-ray and \OIIIF\ luminosities for
  Seyfert 2s derived by Mulchaey et al.\ (1994) and the shaded region shows the 
  3-$\sigma$ error envelope for the sample. It should be noted that
  \OIIIF\ and hard X-ray luminosities are not corrected for
  extinction or absorption in any of these samples.  
The open circle is the data for SMM\,J123549.44+621536.8 
(the only source with precise measurements for both corrections) with absorption 
  corrected \OIIIF\ and hard X-ray luminosities, and the dotted arrow shows 
  the amplitude of these corrections. The bulk of the sources in our
  sample show \OIIIF\ luminosities significantly above those expected
from their X-ray emission assuming the local relationship.  As we show,
applying extinction/absorption corrections to the data will not reduce
this excess (as the correction moves objects parallel to the local
relation). We suggest this excess \OIIIF\ emission arises from shocks
within these galaxies.
}
\end{figure}
\clearpage 

\begin{deluxetable}{lrrrrrl}
\tabletypesize{\scriptsize}
\tablecaption{Observation Log and Basic Parameters}
\tablewidth{0pt}
\tablehead{
\colhead{Object} & 
\colhead{t$_{exp}$\tablenotemark{a}} & 
\colhead{Observation Date} & 
\colhead{Bands} & 
\colhead{L$_{FIR}$\tablenotemark{b}} & 
\colhead{F$_{HX}$\tablenotemark{c}} & 
\colhead{Comment}
}
\startdata
 & & & & & & \\
\multicolumn{7}{c}{Subaru/OHS Observation} \\
 & & & & & & \\
SMMJ123549.44+621536.8 & 7200 & Apr  6 2004 & J\&H &  6.8$^{+1.5}_{-1.1}$\tablenotemark{d} & 12.0$\pm$1.4\tablenotemark{j} &   \\
SMMJ123606.85+621021.4 & 7200 & Feb 15 2005 & J\&H &  8.7$^{+1.8}_{-1.2}$\tablenotemark{d} & 7.4$\pm$1.4\tablenotemark{j} &    \\
SMMJ123622.65+621629.7 & 7200 & Apr  7 2004 & J\&H &  9.0$^{+1.9}_{-2.0}$\tablenotemark{d} & 11.0$\pm$1.4 \tablenotemark{j}&   \\
SMMJ123635.59+621424.1 & 7200 & Feb 16 2005 & J\&H &  7.5$^{+1.5}_{-1.5}$\tablenotemark{d} & 25.0$\pm$1.4\tablenotemark{j} &   \\
SMMJ123716.01+620323.3 & 4000 & Feb 16 2005 & J\&H &  6.3$^{+1.0}_{-1.0}$\tablenotemark{e} & 74.0$\pm$1.4\tablenotemark{j} &   \\
SMMJ123721.87+621035.3 & 7200 & Jun 24 2004 & J\&H &  0.43$^{+0.09}_{-0.09}$\tablenotemark{e} & 21.0$\pm$1.4\tablenotemark{j}&\\
SMMJ131215.27+423900.9 &  600 & Feb 16 2005 & J\&H & 13.9$^{+0.5}_{-2.5}$\tablenotemark{d} & 190.0$\pm$23.0\tablenotemark{k}& \\
SMMJ131222.35+423814.1 & 2000 & Jun 25 2004 & J\&H & 12.7$^{+2.5}_{-2.5}$\tablenotemark{f} & 120.0$\pm$20.0\tablenotemark{k}& \\
SMMJ163639.01+405635.9 & 7200 & Jun 24 2004 & J\&H &  5.5$^{+1.9}_{-1.5}$\tablenotemark{d} & $<$22.0\tablenotemark{l} &   \\
SMMJ163650.43+405734.5 & 7000 & May 18 2003 & J\&H & 50.5$^{+15.0}_{-15.8}$\tablenotemark{d} & $<$22.0\tablenotemark{l} & Smail et al. (2003)\\
MMJ163655+4059         & 3600 & Apr  7 2004 & J\&H & 10.9$^{+2.2}_{-3.8}$\tablenotemark{d} & 150.0$\pm$21.0\tablenotemark{l} & \\
SMMJ163706.51+405313.8 & 7200 & Apr  6 2004 & J\&H &  7.2$^{+4.2}_{-3.0}$\tablenotemark{d} &  $<$22.0\tablenotemark{l} &      \\
SMMJ221733.02+000906.0 & 3000 & Jun 24 2004 & J\&H &  1.9$^{+0.4}_{-0.4}$\tablenotemark{f} & $<$28.0\tablenotemark{m} &  \\
SMMJ221733.79+001402.1 & 7200 & Jun 25 2004 & J\&H &  4.9$^{+1.9}_{-2.6}$\tablenotemark{d} & $<$28.0\tablenotemark{m} & \\
 & & & & & & \\
\multicolumn{7}{c}{VLT/ISAAC Observation} \\
 & & & & & & \\
SMMJ02399-0134         & 4500 & Nov 24,25 2004 & J &  6.5$^{+1.3}_{-1.3}$\tablenotemark{h} & 32.0$\pm$5.0\tablenotemark{n} &   \\
                       & 2400 & Nov 25 2004 &   z  &                                       &                               &   \\
SMMJ030227.73+000653.5 & 6000 & Nov 23 2004 & J    &  5.8$^{+2.4}_{-0.8}$\tablenotemark{d} & 60.0$\pm$10.0\tablenotemark{o} &   \\
RGJ030257.94+001016.3  & 4500 & Nov 23 2004 & H    &  7.7$^{+1.4}_{-1.4}$\tablenotemark{d} & 60.0$\pm$10.0\tablenotemark{o} &   \\
SMMJ105702.50-033602.6 & 4500 & Nov 23 2004 & H    &  5.0$^{+1.0}_{-1.0}$\tablenotemark{i} & $<$10.0\tablenotemark{i} &   \\
SMMJ221737.39+001025.1 & 9000 & Nov 23,24 2004 &K  & 21.0$^{+4.2}_{-4.2}$\tablenotemark{f} & $<$28.0\tablenotemark{m} &   \\
                       & 6000 & Nov 25 2004 & H    &                                       &                          &   \\
 & & & & & & \\
\multicolumn{7}{c}{Keck/NIRSPEC Observation}\\
 & & & & & &\\
SMMJ09431+4700(H6)     & 2400 & Apr  8 2004 & K    & 15.0$^{+3.0}_{-3.0}$\tablenotemark{g} & $<$13.0\tablenotemark{g} &  \\
SMMJ09431+4700(H7)     & 2400 & Apr  8 2004 & K    & 15.0$^{+3.0}_{-3.0}$\tablenotemark{g} & $<$13.0\tablenotemark{g} &  \\
SMMJ131201.17+424208.1 & 2400 & Apr  8 2004 & K    & 20.2$^{+4.0}_{-4.0}$\tablenotemark{f} & 52.8$\pm$40.0\tablenotemark{k} & \\
 & & & & & & \\
\multicolumn{7}{c}{From literature}\\
 & & & & & & \\
SMMJ04431+0210         &      &             &      & 3.5$^{+1.5}_{-1.5}$\tablenotemark{h} & --- & Frayer et al. (2003)\\
SMMJ14011+0252(J1)     &      &             &      & 6.8$^{+2.5}_{-2.5}$\tablenotemark{h} & $<$44.0\tablenotemark{p} & Motohara et al.(2005)\\
SMMJ163658.19+410523.8 &      &             &      & 10.9$^{+2.2}_{-3.8}$\tablenotemark{d} & $<$22.0\tablenotemark{l} & Simpson et al.(2004)\\
\enddata

\tablenotetext{a}{in seconds.}
\tablenotetext{b}{FIR luminosity in 10$^{12} \lsun$}
\tablenotetext{c}{Hard X-ray flux in 10$^{16}$ erg sec$^{-1}$ cm$^{-2}$}
\tablenotetext{d}{Swinbank et al. (2004)}
\tablenotetext{e}{Alexander et al. (2005b)}
\tablenotetext{f}{Smail et al. (2004)}
\tablenotetext{g}{Ledlow et al. (2002)}
\tablenotetext{h}{Smail et al. (2002)}
\tablenotetext{i}{van Dokkum et al. (2004)}
\tablenotetext{j}{Alexander et al. (2003)}
\tablenotetext{k}{Mushotzky et al. (2000)}
\tablenotetext{l}{Manners et al. (2003)}
\tablenotetext{m}{Basu-Zych \& Scharf (2005)}
\tablenotetext{n}{Bautz et al. (2000)}
\tablenotetext{o}{Waskett et al. (2004)}
\tablenotetext{p}{Fabian et al. (2000)}

\end{deluxetable}

\clearpage 

\begin{turnpage}
\begin{table}
\begin{center}
\caption{Summary of results}
\begin{tabular}{lrrrrrrcccrl}
\tableline
Object & 
$z$\tablenotemark{a} & 
H$\alpha$ flux & 
H$\beta$ flux & 
[O III]$_{\lambda5007}$ flux & 
[O III]$_{\lambda4959}$ flux & 
[O II]$_{\lambda3727}$ flux & 
\multicolumn{3}{c}{Class} & 
rest FWHM$_{H_{\beta}}$ & 
Comment \\
     &
     &
\multicolumn{5}{c}{10$^{-16}$erg cm$^{-2}$ sec$^{-1}$} & 
UV & 
H$_\alpha$ & 
Opt   & 
km$\cdot$sec$^{-1}$  &
\\    
\tableline
SMMJ02399-0134          & 1.061\tablenotemark{c} &  75.8$\pm$ 15.0&    $<$11.0     &  $<$11.0       &  $<$11.0       & ---          & AGN & AGN & ---  & 1530$\pm$500\tablenotemark{b} \\
SMMJ030227.73+000653.5  & 1.408                  &  15.2$\pm$ 2.0 &   1.9$\pm$ 1.1 &  10.7$\pm$ 3.2 &   7.2$\pm$ 2.6 & ---          & SB & AGN & AGN   & $<$100        & \\
RGJ030258.94+001016.3   & 2.239                  &   1.8$\pm$ 0.5 &   $<$0.3       & 9.2$\pm$ 2.1 &   0.85$\pm$ 0.30 & ---          & int & AGN & AGN  & ---           & \\ 
SMMJ09431+4700(H6)      & 3.350                  & ---            &   0.3$\pm$ 0.2 &   1.7$\pm$ 0.3 &   0.7$\pm$ 0.4 & ---          & SB & --- &SB?   & 350$\pm$50    & Extended \OIIIF\ \\
SMMJ09431+4700(H7)      & 3.347                  & ---            & $<$0.15        &   0.5$\pm$ 0.1 & $<$0.15        & ---          & --- & --- & ---  & ---           & \\
SMMJ105702.50-033602.6  & 2.423\tablenotemark{b} &   0.6$\pm$  0.2&   $<$0.19      &   $<$0.19      &   $<$0.19      & ---          & --- & SB & ---   & ---           & van Dokkum et al. (2004); miss the slit? \\
SMMJ123549.44+621536.8  & 2.195                  &  15.0$\pm$ 1.0 &   1.6$\pm$ 1.0 &   6.4$\pm$ 1.3 &   1.9$\pm$ 0.8 & 2.9$\pm$ 0.8 & SB & int & AGN   & 2150$\pm$500  & \OIIIF\ double peaks and extended  \\
SMMJ123606.85+621021.4  & 2.505\tablenotemark{b} &   2.0$\pm$ 0.3 &   $<$0.043     &   $<$0.043     &   $<$0.043     & $<$0.043     & SB & int & ---   & ---           & slit on companion object?\\
SMMJ123622.65+621629.7  & 2.462\tablenotemark{c} &   3.4$\pm$ 0.6 &   $<$0.8       &   $<$0.8       &   $<$0.8       & 2.1$\pm$ 0.5 &  SB & SB & ---   & ---           & \\
SMMJ123635.59+621424.1  & 2.005                  &  11.1$\pm$ 1.2 & ---            &   2.0$\pm$ 0.6 &   2.4$\pm$ 0.5 & ---          & AGN & AGN & AGN? & ---           & \\
SMMJ123716.01+620323.3  & 2.053                  &  ---           &  28.1$\pm$ 3.2 &  19.5$\pm$ 2.4 & 9.7$\pm$ 2.2   & ---          & QSO & --- & AGN  & 2130$\pm$500  & multiple peaks and extended \OIIIF\ \\
SMMJ123721.87+621035.3  & 0.979\tablenotemark{b} &   6.2$\pm$  1.1& ---            & ---            & ---            & ---          & --- & SB & ---   & ---           & \\
SMMJ131201.17+424208.1  & 3.408                  & ---            & 0.18$\pm$0.15  & 0.79$\pm$0.19  & 0.27$\pm$0.19  & ---          & AGN & --- & AGN? & ---           & \\
SMMJ131215.27+423900.9  & 2.555                  &  11.8$\pm$ 1.0 &   3.1$\pm$ 2.6 &   5.6$\pm$ 2.2 &   2.2$\pm$ 1.6 & $<$0.23          & QSO & --- & AGN  & 2540$\pm$500  & \\
SMMJ131222.35+423814.1  & 2.560                  & ---            &  12.1$\pm$ 4.5 &   5.5$\pm$ 2.5 &   2.6$\pm$ 2.5 & 2.0$\pm$ 1.5 & QSO & --- & AGN  & 2580$\pm$1000 & \\
SMMJ163639.01+405635.9  & 1.485\tablenotemark{c} &   7.3$\pm$ 0.7 &   0.7$\pm$ 0.5 &   2.5$\pm$ 0.5 &   1.1$\pm$ 0.4 & ---          & SB & SB & int    & $<$1400       & Extended \hal\    \\
SMMJ163650.43+405734.5  & 2.380                  &  14.2$\pm$ 1.5 &   2.3$\pm$ 0.6 &  27.0$\pm$ 2.1 &   6.8$\pm$ 1.3 & 14.0$\pm$ 0.8&  int & AGN & AGN & 3720$\pm$500  & Reevaluated after Smail et al. (2003)\\
MMJ163655+4059          & 2.605                  &  18.4$\pm$ 2.4 &   2.7$\pm$ 1.0 &  47.1$\pm$ 1.6 &  15.7$\pm$ 1.6 & 3.6$\pm$ 0.9 & AGN & AGN & AGN  & 2410$\pm$600  & Blue wings in \hb\ and \OIIIE\ \\
SMMJ163706.51+405313.8  & 2.373                  &   7.4$\pm$ 1.4 &   0.6$\pm$ 0.3 &   5.7$\pm$ 0.6 &   1.9$\pm$ 0.6 & 1.0$\pm$ 0.3 & AGN & AGN & AGN  & 1590$\pm$500  & \\
SMMJ221733.02+000906.0  & 0.926\tablenotemark{b} &   9.6$\pm$ 1.2 & ---            & ---            & ---            & ---          & --- & SB & ---   & ---           & \\
SMMJ221733.79+001402.1  & 2.551                  &   8.5$\pm$ 3.5 &   1.2$\pm$ 0.6 &   1.2$\pm$ 0.2 &   $<$0.5       & 0.7$\pm$ 0.3 & SB & SB & AGN?   & 1860$\pm$600  & \\
SMMJ221737.39+001025.1  & 2.610\tablenotemark{b} &  20.7$\pm$ 6.0 &   0.6 $\pm$0.4 &   6.0$\pm$ 0.4 &   1.2$\pm$ 0.4 & ---          &  SB & AGN & AGN  & 290$\pm$50    & \\
                        &                        &                &                &                &                &              &   &   &  \\
SMMJ04431+0210          & 2.510                  &   1.6$\pm$ 0.1 &   $<$0.3       &   0.4$\pm$ 0.1 & ---            & ---          & --- & AGN? & AGN & ---           & Frayer et al. (2003)\\
SMMJ14011+0252(J1)      & 2.565                  &   1.3$\pm$ 0.4 &   0.3$\pm$ 0.1 &   $<$0.2 & $<$0.2\tablenotemark{d}&0.4$\pm$0.1 & SB & SB & SB     & ---           & Motohara et al. (2005) \\
SMMJ163658.19+410523.8  & 2.448                  &   1.9$\pm$ 0.4 &   0.2$\pm$ 0.1 &   0.4$\pm$ 0.1 &   0.3$\pm$ 0.1 & 0.4$\pm$ 0.2 & SB & SB & AGN    & ---           &  Simpson et al. (2004)\\
                        &                        &                &                &                &                &              &  \\
Composite(all)        &    &   & 1.0 & 0.7$^{+0.1}_{-0.1}$ &  0.4$^{+0.1}_{-0.1}$ & 0.3$^{+0.2}_{-0.1}$ &  & & & 3100$\pm$500  &  \\
Composite(QSO)        &    &   & 1.0 & 0.36$^{+0.33}_{-0.18}$ & 0.19$^{+0.21}_{-0.11}$  & $<$0.12  &  & & & 3200$\pm$1000  &  \\
Composite(OPT-AGN)    &    &   & 1.0 & 3.2$^{+1.0}_{-0.6}$  & 1.2$^{+0.5}_{-0.3}$ & 1.2$^{+0.7}_{-0.4}$ & & & & 1730$\pm$500 &  \\
\tableline
\end{tabular}

\tablenotetext{a}{based on \OIIIF\ line measurement.}
\tablenotetext{b}{based on \hal\ line measurement.}
\tablenotetext{c}{based on \OII\ line measurement.}
\tablenotetext{d}{detected in Tecza et al. 2004}
\tablecomments{}
\end{center}
\end{table}

\end{turnpage}

\clearpage

\end{document}